\documentclass[12pt]{article}
\usepackage{epsf}
\usepackage{cite}

\newcommand{\h}{\bar{h}}
\def\theequation{\arabic{section}.\arabic{equation}}

\begin{document}
~\vspace{-1.cm}
\begin{flushright}
MC-TH-2002-12\\[-0.15cm]
SINP/TNP/02-36\\[-0.15cm]
WUE-ITP-2002-037\\[-0.15cm] 
hep-ph/0212169\\[-0.15cm]
December 2002
\end{flushright}   

\begin{center}
{\Large {\bf Neutrinoless Double Beta Decay }}\\[0.3cm]
{\Large {\bf from Singlet Neutrinos in Extra Dimensions}}\\[1.5cm] 
\hspace{-0.2cm}{\large 
G. Bhattacharyya$^{\, a}$, H.V. Klapdor-Kleingrothaus$^{\, b}$, 
H. P\"as$^{\, c}$, A. Pilaftsis$^{\, d}$}\\[0.7cm]
$^a${\it Saha Institute of Nuclear Physics, 1/AF Bidhan Nagar,
         Kolkata 700064, India}\\[0.2cm] 
$^b${\it Max-Planck-Institut f\"ur Kernphysik, P.O. Box 103980,
         D-69029 Heidelberg, Germany}\\[0.2cm]
$^c${\it Institut f\"ur Theoretische Physik und Astrophysik, 
         Universit\"at W\"urzburg,\\ Am Hubland, 97074
         W\"urzburg, Germany}\\[0.2cm] 
$^d${\it Department of Physics and Astronomy, University of Manchester,\\ 
         Manchester M13 9PL, United Kingdom}
\end{center}
\vskip1.cm   \centerline{\bf   ABSTRACT}    {\small   
We  study  the  model-building   conditions  under  which  a  sizeable
$0\nu\beta\beta$-decay signal to the recently reported level of~0.4~eV
is due to Kaluza--Klein singlet neutrinos in theories with large extra
dimensions.  Our  analysis is based  on 5-dimensional singlet-neutrino
models   compactified   on    an   $S^1/Z_2$   orbifold,   where   the
Standard--Model fields  are localized  on a 3-brane.   We show  that a
successful  interpretation  of  a  positive signal  within  the  above
minimal 5-dimensional framework would require a non-vanishing shift of
the 3-brane from  the orbifold fixed points by  an amount smaller than
the  typical scale (100~MeV)$^{-1}$  characterizing the  Fermi nuclear
momentum.   The  resulting  5-dimensional  models predict  a  sizeable
effective  Majorana-neutrino  mass that  could  be  several orders  of
magnitude larger than the  light neutrino masses.  Most interestingly,
the  brane-shifted models  with only  one bulk  sterile  neutrino also
predict novel  trigonometric textures  leading to mass  scenarios with
hierarchical  active  neutrinos  and  large  $\nu_\mu$-$\nu_\tau$  and
$\nu_e$-$\nu_\mu$   mixings  that  can   fully  explain   the  current
atmospheric and solar neutrino data. }

\newpage

\setcounter{equation}{0}
\section{Introduction}

Recently,   realizations of  phenomenologically  viable  theories with
large   compact  dimensions   of~TeV size~\cite{extra}  have  enriched
dramatically  our perspectives in  searching   for physics beyond  the
Standard  Model   (SM).     Among   the  possible   higher-dimensional
realizations,  sterile    neutrinos  propagating   in     large  extra
dimensions~\cite{DDG2,ADDM,DS,AP1,IP,nuextra,LRRR,DLP}    may  provide
interesting  alternatives for generating  the observed  light neutrino
masses.  On the other hand, detailed  experimental studies of neutrino
properties may even shed light on the geometry and/or shape of the new
dimensions.  In this context, one  of the most sensitive  experimental
approaches to neutrino  masses and their  properties is the search for
neutrinoless double  beta decay~\cite{doi85}. Neutrinoless double beta
decay, denoted in short as $0\nu\beta\beta$, corresponds to two single
beta  decays~\cite{Klapdor,HKK}    occurring simultaneously    in  one
nucleus,  thereby  converting  a    nucleus  $(Z,A)$  into  a  nucleus
$(Z+2,A)$, i.e.\
\begin{displaymath}        
^{A}_{Z}\,X\ \to\ ^A_{Z+2}\,X\: +\: 2 e^-\; . 
\end{displaymath}
This process   violates  lepton number by   two  units  and  hence its
observation would  signal physics  beyond  the  SM.  To a   very  good
approximation, the half life  for a $0\nu\beta\beta$ decay mediated by
light neutrinos is given by
\begin{equation}
\label{t1/2}
[T^{0\nu\beta\beta}_{1/2}]^{-1}\ =\ \frac{|\langle m \rangle |^2}{m^2_e}\
|{\cal M}_{0\nu\beta\beta}|^2\, G_{01}\; , 
\end{equation} 
where $\langle  m  \rangle$ denotes  the effective  neutrino  Majorana
mass, $m_e$  is the electron mass  and ${\cal M}_{0\nu\beta\beta}$ and
$G_{01}$ denote the appropriate 
nuclear matrix element and the phase space factor,
respectively.   For    details, see~\cite{doi85,Klapdor,HKK}  and  our
discussion in Section~4.
 
Most recently, the Heidelberg--Moscow collaboration has reanalyzed its
experimental data~\cite{HMexp},  using appropriate statistical methods
as  well  as  new  information  from  the  form  of  the  contributing
background.   They found  an excess  of $0\nu\beta\beta$  events, with
statistical  significance 2.2--3.1~$\sigma$  depending  on the  method
used.   {}From this result,  a half-life  of 1.5$^{+16.8}_{-0.7}\times
10^{25}$ years at 95\% confidence level (CL) for $^{76}$Ge is deduced,
which implies  an absolute  value for the  effective Majorana-neutrino
mass:
\begin{equation}
  \label{HMexp}
|\langle m \rangle|\ =\ 0.39^{+0.45}_{-0.34}~{\rm eV}\quad (95\%~{\rm CL})\; ,
\end{equation}
allowing  an uncertainty of the  nuclear matrix element values of $\pm
50\%$. 

The above experimental result~(\ref{HMexp}), combined with information
from solar  and  atmospheric neutrino  data, restricts   the admissible
forms of the light-neutrino mass  hierarchies in 4-dimensional  models
with 3 left-handed  (active) neutrinos.  The allowed scenarios contain
either degenerate neutrinos  or  neutrinos that have an   inverse mass
hierarchy~\cite{KS}.   Evidently,  a   successful  interpretation of a
positive  $0\nu\beta\beta$  signal of  the  appropriate size mentioned
above imposes certain constraints on the structure of a theory.  Here,
we  study  these  constraints   on  the  model  building   of  minimal
5-dimensional theories compactified  on a $S^1/Z_2$ orbifold.   Within
the  framework of  theories   with large   extra dimensions,  previous
studies on neutrinoless  double beta decays were performed  within the
context  of   higher-dimensional models   that  utilize  the   shining
mechanism from a distant brane~\cite{MLP} and of theories with wrapped
geometric       space~\cite{Huber}.         In   Ref.~\cite{MLP},  the
$0\nu\beta\beta$ decay  is   accompanied with emission    of Majorons,
whereas the prediction in~\cite{Huber}  falls  short by two  orders of
magnitude to account for the observable excess in~(\ref{HMexp}).

In this paper, we   consider an even more  minimal  higher-dimensional
framework of lepton-number violation, in  which only one 5-dimensional
(bulk) sterile neutrino is added  to the field content of  the SM.  In
this minimal model, the SM   fields are localized on a   4-dimensional
Minkowski subspace, also termed 3-brane.   The violation of the lepton
number  may occur in three distinct  ways: (i)~by adding lepton-number
violating bilinears of the Majorana   type in the Lagrangian;  (ii)~by
generating    lepton-number-violating      mass  terms     through the
Scherk--Schwartz mechanism~\cite{SS}; (iii)~by simultaneously coupling
the    $Z_2$-even    and  $Z_2$-odd     two-component  spinors of  the
5-dimensional sterile neutrino  to the same left-handed charged lepton
state.   As we  will see in  Section~2,   the last case~(iii)  is only
possible  if the  3-brane describing  our observable  world is shifted
from the  $S^1/Z_2$ orbifold fixed point.  Here,  we  should also note
that after   integration of the    extra dimension, the  5-dimensional
orbifold model    predicts  an infinite tower   of  Kaluza--Klein (KK)
neutrinos, for which the cases (i) and (ii) become fully equivalent.

One salient feature of the $S^1/Z_2$ orbifold compactification is that
the KK neutrinos  group themselves into approximately degenerate pairs
of opposite  CP  parities.  As  a  result, the lepton-number-violating
KK-neutrino   effects cancel  each   other    and  so  the   predicted
$0\nu\beta\beta$  decay turns out  to  be exceedingly small to account
for the recent  observable excess.  The latter appears  to be a  major
obstacle in theories with large extra dimensions and imposes by itself
constraints  on the model-building  of higher-dimensional theories.  A
minimal way  that avoids the   above disastrous CP-parity cancellation
effects on the $0\nu\beta\beta$  decay  amplitude would be  to arrange
the opposite CP-parity KK neutrinos  to  couple to the $W^\pm$  bosons
with   unequal strength.   Within  the  minimal 5-dimensional orbifold
model outlined above, such  a realization can  be accomplished only if
the  3-brane is  displaced  from one of  the  $S^1/Z_2$ orbifold fixed
points.  In our phenomenological   bottom-up approach, the amount   of
brane-shifting  is not arbitrary but dictated  by the requirement that
the  model can accommodate  the result~(\ref{HMexp}) for the effective
Majorana-neutrino  mass.  In particular, we will  see in Section~4 how
the  resulting  brane-shifted    5-dimensional models can  predict   a
sizeable effective Majorana-neutrino mass that could be several orders
of magnitude larger than the light neutrino masses  and hence than the
difference of  their squares  as  required from  neutrino  oscillation
data.

Another  important constraint on   the structure of higher-dimensional
neutrino theories  arises from their ability  to explain the solar and
atmospheric   neutrino data by    means of neutrino oscillations.   In
particular,     orbifold  models with one   bulk    neutrino, as those
considered earlier in  the literature~\cite{DDG2,DS,nuextra,LRRR,DLP},
seem     to         prefer the    Small       Mixing      Angle  (SMA)
Mikheev-Smirnov-Wolfenstein (MSW)  solution~\cite{MSW} which is highly
disfavoured by recent  neutrino data analyses.  Alternatively, if  all
neutrino data are to be  explained by oscillations of active neutrinos
with  a   small    admixture  of sterile    KK    component,  then the
compactification scale has to be much higher than the brane-Dirac mass
terms.  After integrating  out  the bulk  neutrino of  the model,  the
effective light-neutrino mass matrix has  a rather restricted form; it
is  effectively of rank 1.  As  a result, two out  of the three active
neutrinos are massless.  This is  rather  undesirable, since only  one
neutrino-mass difference can be  formed in this case, so accommodating
all       neutrino    oscillation         data        proves    rather
problematic~\cite{nuextra,LRRR,DLP}.    However,   the earlier studies
have not  included  the   possibility of   a shifted  brane.    As was
mentioned above, brane-shifting  gives rise to sizeable  lepton-number
violation. Hence, the tree-level rank-1 form of the effective neutrino
mass  matrix  can  be  significantly  modified   through lepton-number
violating Yukawa  terms.  As we will see  in  Section 5, the resulting
neutrino  mass matrix has   sufficiently   rich structure  to   enable
adequate description of the neutrino data.

Our paper is organized as  follows: Section~2 describes the low-energy
structure of the  5-dimensional orbifold models. Technical details are
relegated   to    the appendices.   In   Section~3,    we  study   the
renormalization-group (RG) effects  of  the neutrino Yukawa  couplings
and their possible impact on the $0\nu\beta\beta$ decay amplitude.  In
Section~4 we give estimates of  the effective Majorana-neutrino  mass,
which are predicted  in   these models presented  in   Section~2.   In
Section 5, we discuss the compatibility  of such models with solar and
atmospheric neutrino  data.  Finally,   we  draw our   conclusions  in
Section~6.

\setcounter{equation}{0}
\section{Minimal higher-dimensional neutrino models}

In  this section, we will describe  the  basic low-energy structure of
minimal higher-dimensional extensions of the  SM that include  singlet
neutrinos.   In  particular, we assume    that singlet neutrinos being
neutral under   the SU(2)$_L\otimes$U(1)$_Y$ gauge  group  can  freely
propagate   in  a    higher-dimensional   space  of   $[1+(3+\delta)]$
dimensions, the so-called bulk, whereas all SM particles are localized
in  a $(1+3)$-dimensional subspace, known  as 3-brane or simply brane.
However,  even singlet neutrinos themselves may  live in a subspace of
an even  higher-dimensional  space of $[1+(3+n_g)]$  dimensions,  with
$\delta \le n_g$, in which gravity propagates.

We shall restrict  our study to 5-dimensional  models, i.e.\  the case
$\delta  =  1$,  where the singlet   neutrinos  are compactified on  a
$S^1/Z_2$  orbifold.   Specifically,  the   leptonic  sector  of   our
5-dimensional model consists of the SM lepton fields:
\begin{equation}
  \label{LSM}
L(x)\ =\ \left( \begin{array}{c} \nu_l (x) \\ l_L (x) \end{array}
\right) ,\qquad
l_R (x)\,,
\end{equation}
with $l = e,\mu,\tau$, and one 5-dimensional (bulk) singlet neutrino:
\begin{equation}
  \label{Nu}
 N(x,y)\ =\ \left( \begin{array}{c} \xi (x,y) \\ 
\bar{\eta} (x,y) \end{array} \right)\, ,
\end{equation}
where  $y$ denotes  the  additional compact dimension,  and $\xi$  and
$\eta$ are 5-dimensional two-component   spinors.  The SM  leptons are
localized at  the  one  of the  two  fixed  points of   the  $S^1/Z_2$
orbifold, e.g.\ $y = 0$. For generality, we will assume that the brane
is shifted from the orbifold fixed point to $y = a$.

As   usual,  we  impose the   periodic   boundary condition $N(x,y)  =
N(x,y+2\pi R)$ with  respect to $y$ dimension  on the singlet neutrino
field.  In  addition,  the  action  of  $S^1/Z_2$ orbifolding  on  the
5-dimensional spinors  $\xi$    and $\eta$   entails  the   additional
identifications:
\begin{equation}
  \label{yparity}
\xi (x,y)\ =\ \xi (x,-y)\,,\qquad  \eta (x,y)\ =\ - \eta (x,-y)\,.
\end{equation}
In other   words,  the spinors  $\xi$ and   $\eta$ are   symmetric and
antisymmetric under a $y$ reflection, respectively.

With the above  definitions, the most generic  effective 4-dimensional
Lagrangian       of        such      a      model          is    given
by~\cite{DDG2,AP1}\footnote{Further non-covariant    extensions  to
this model have been considered in~\cite{LRRR}.}
\begin{eqnarray}
  \label{Leff} 
{\cal L}_{\rm  eff} & =& \int\limits_0^{2\pi R}\!\! dy\
\bigg\{\,   \bar{N} \Big(   i\gamma^\mu  \partial_\mu\,  +\,  \gamma_5
\partial_y \Big) N\ -\  \frac{1}{2}\,\Big( M N^T  C^{(5)-1} N\ +\ {\rm
h.c.}    \Big)         \nonumber\\     
&&+\,\delta   (y-a)\,  \bigg[\, \frac{{h}^l_1}{(M_F)^{\delta/2}}\, 
L\tilde{\Phi}^*      \xi\,    +\, \frac{{h}^l_2}{(M_F)^{\delta/2}}\,  L 
\tilde{\Phi}^*   \eta\ +\   {\rm h.c.}\,\bigg]\ +\ 
\delta (y-a)\, {\cal L}_{\rm SM}\, \bigg\}\, ,\quad
\end{eqnarray}
where $\tilde{\Phi} =  i\sigma_2 \Phi^*$ is the  hypercharge-conjugate
of the SM Higgs  doublet $\Phi$, with hypercharge  $Y(\Phi) = 1$,  and
${\cal L}_{\rm SM}$ denotes the SM Lagrangian which is restricted on a
brane at $y=a$~\cite{DDG2}.  In   addition, $M_F$ is   the fundamental
$n_g$-dimensional Planck scale and $\delta =  1$ for sterile neutrinos
propagating  in 5 dimensions.  Notice  that the mass term $m_D \bar{N}
N$ is not allowed  in~(\ref{Leff}), as a result  of the $Z_2$ discrete
symmetry.  Finally,  in   writing~(\ref{Leff}),   we   have used   the
following conventions:
\begin{equation}
  \label{gammas}
\gamma^\mu\ =\ \left( \begin{array}{cc} 0 & \sigma^\mu \\
              \bar{\sigma}^\mu & 0 \end{array} \right)\,,\quad
\gamma_5\ =\ \left(\!
\begin{array}{cc} -{\bf 1}_2 & 0 \\
  0 & {\bf 1}_2 \end{array} \!\right)\, , \quad
C^{(5)}\ =\ - \gamma_1 \gamma_3\ =\ \left(\!
\begin{array}{cc} -i\sigma_2 & 0 \\
  0 & -i\sigma_2 \end{array} \!\right)\, ,
\end{equation}
with $\sigma^\mu =    ({\bf  1}_2, \mbox{\boldmath $\sigma$} )$    and
$\bar{\sigma}^\mu = ({\bf 1}_2, -  \mbox{\boldmath $\sigma$} )$, where
$\sigma_{1,2,3}$ are the usual Pauli matrices.

We now proceed with  the compactification of  the $y$ dimension of the
$S^1/Z_2$     orbifold  model.  Because    of   their   symmetric  and
antisymmetric   properties~(\ref{yparity}) under   $y$ reflection, the
two-component spinors $\xi$ and $\eta$   can be expanded in a  Fourier
series of cosine and sine harmonics:
\begin{eqnarray}
  \label{xi}
\xi (x,y) &=& \frac{1}{\sqrt{2\pi R}}\ \xi_0 (x)\ +\ 
\frac{1}{\sqrt{\pi R}}\ \sum_{n=1}^\infty\, \xi_n (x)\ 
                                       \cos\bigg(\,\frac{ny}{R}\,\bigg)\,,\\
  \label{eta}
\eta (x,y) & =& \frac{1}{\sqrt{\pi R}}\ \sum_{n=1}^\infty\, \eta_n (x)\ 
                                       \sin\bigg(\,\frac{ny}{R}\,\bigg)\,,
\end{eqnarray}
where the chiral spinors $\xi_n (x)$ and $\eta_n (x)$ form an infinite
tower of   KK  modes. 

After substituting~(\ref{xi})  into  (\ref{Leff}) and  integrating out
the $y$ coordinate, we obtain the effective 4-dimensional Lagrangian
\begin{eqnarray}
  \label{Leff1KK}
{\cal L}_{\rm eff} & = & {\cal L}_{\rm SM}\ +\ \bar{\xi}_0
( i\bar{\sigma}^\mu \partial_\mu) \xi_0\ 
+\ \Big(\, \bar{h}^{l(0)}_1\, L\tilde{\Phi}^* \xi_0\ -\
\frac{1}{2}\, M\, \xi_0\xi_0\ +\ {\rm h.c.}\,\Big)\
 +\ \sum_{n=1}^\infty\, \bigg[\, \bar{\xi}_n 
( i\bar{\sigma}^\mu \partial_\mu) \xi_n\nonumber\\
&& +\, \bar{\eta}_n ( i\bar{\sigma}^\mu \partial_\mu) \eta_n\
+\ \frac{n}{R}\, \Big( \xi_n \eta_n\, +\, \bar{\xi}_n
\bar{\eta}_n\Big) -\ \frac{1}{2}\, M\, 
\Big( \xi_n\xi_n\, +\, \bar{\eta}_n\bar{\eta}_n\
+\ {\rm h.c.}\Big)\nonumber\\
&& +\, \sqrt{2}\, \Big(\, \bar{h}^{l(n)}_1\, L\tilde{\Phi}^* \xi_n\ +\
\bar{h}^{l(n)}_2\, L\tilde{\Phi}^* \eta_n\ +\ {\rm h.c.}\,\Big)\, \bigg]\, ,
\end{eqnarray}
where 
\begin{eqnarray}
  \label{h1n}
\bar{h}^{l(n)}_1 &=& \frac{h^l_1}{(2\pi M_F R)^{\delta/2}}\ 
\cos \bigg(\,\frac{na}{R}\,\bigg)\ =\ \bigg(\,\frac{M_F}{M_{\rm
P}}\,\bigg)^{\delta/n_g}\  
h^l_1 \cos \bigg(\,\frac{na}{R}\,\bigg)\,,\\
  \label{h2n}
\bar{h}^{l(n)}_2 &=& \frac{h^l_2}{(2\pi M_F R)^{\delta/2}}\ 
\sin \bigg(\,\frac{na}{R}\,\bigg)\ =\ \bigg(\,\frac{M_F}{M_{\rm
P}}\,\bigg)^{\delta/n_g}\  
h^l_2 \sin \bigg(\,\frac{na}{R}\,\bigg)\, . 
\end{eqnarray}
In deriving the last step on the RHS's of (\ref{h1n}) and (\ref{h2n}),
we have employed the basic relation among the Planck mass $M_{\rm P}$,
the corresponding $n_g$-dimensional     Planck mass   $M_F$  and   the
compactification radii $R$ (all taken to be of equal size):
\begin{equation}
  \label{MF}
M_{\rm P}\ =\ (2\pi\, M_F\, R)^{n_g/2}\, M_F\ .
\end{equation}
{}From~(\ref{h1n})   and  (\ref{h2n}),  we    see  that  the   reduced
4-dimensional Yukawa couplings $\bar{h}^{(n)}_{1,2}$ can be suppressed
by   many orders of  magnitude\cite{ADDM,DDG2}  if  there  is  a large
hierarchy between $M_{\rm  P}$  and the quantum  gravity  scale $M_F$.
Thus, if  gravity  and bulk neutrinos feel  the  same number of  extra
dimensions, i.e.\ $\delta  = n_g$, the 4-dimensional  Yukawa couplings
$\bar{h}^{(n)}_1$ and $\bar{h}^{(n)}_2$  are naturally suppressed by a
huge  factor $M_F/M_{\rm P} \sim 10^{-15}$,  for $M_F \approx 10$~TeV.
{}From~(\ref{Nu}), we  observe  that $\xi$ and $\bar{\eta}$  belong to
the same multiplet   and hence have   the same lepton number. It  then
follows   from  (\ref{Leff1KK})  that  the  simultaneous  presence  of
$\bar{h}^{(n)}_1$ and $\bar{h}^{(n)}_2$ in  an amplitude gives rise to
lepton number violation by two units.

We should note  that the  above  large suppression factor  can be also
obtained  in  a 5-dimensional neutrino  model   ($\delta = 1$),  where
gravity propagates  in  a  6-dimensional space with   compactification
radii $R_1$ and $R_2$ of unequal size ($n_g = 2$). In this case, one
has to use 
the general toroidal compactification condition:
\begin{equation}
  \label{MFgen}
M_{\rm P}\ =\ (2\pi\, M_F)^{n_g/2}\, (R_1 R_2\dots R_{n_g})^{1/2}\,
M_F\ .
\end{equation}
Note that (\ref{MFgen}) reduces  to (\ref{MF}) if all compactification
radii are equal. With the help of (\ref{MFgen}), we find for $n_g =2$
\begin{equation}
  \label{hnew}
\frac{h^l_{1,2}}{(2\pi M_F R_1)^{1/2}}\ =\ (2\pi M_F R_2 )^{1/2}\
\frac{M_F}{M_{\rm P}}\, h^l_{1,2}\, .
\end{equation}
We easily see that if $R_2 \sim  1/M_F$, the original Yukawa couplings
$h^l_{1,2}$ undergo the  same large degree  of suppression by a factor
$M_F/M_{\rm P}$.

If the brane were located at the one of the two orbifold fixed points,
e.g.\ at $y=0$, the operator  $L\tilde{\Phi}^*\eta$ would be absent as
a consequence  of the $Z_2$ discrete symmetry.   However, if the brane
is shifted  by an amount   $a\ne 0$, the  above  operator is no longer
absent. In fact, as we will  see in Section 4,  the coexistence of the
two    operators $L\tilde{\Phi}^*\xi$ and $L\tilde{\Phi}^*\eta$ breaks
the  lepton  number leading  to   observable effects  in  neutrinoless
double beta decay experiments.

Let us now introduce the weak basis for the KK-Weyl spinors
\begin{equation}
  \label{xieta}
\chi_{\pm n}\ =\ \frac{1}{\sqrt{2}}\, (\,\xi_n\: \pm\:
\eta_n\,)\,,
\end{equation}
which enables  to express the  effective kinetic term of  the neutrino
sector as follows:
\begin{eqnarray}
  \label{Lkin1}
{\cal L}_{\rm kin} &=& \bar{\chi}\,i\bar{\sigma}^\mu\partial_\mu\,\chi\:
-\: \bigg(\,\frac{1}{2}\, \chi^T\, {\cal M}\, \chi\quad +\quad 
{\rm h.c.}\bigg)\, ,
\end{eqnarray}
where   $\chi^T = \big(\nu_l,\  \xi_0,\ \chi_1,\ \chi_{-1},\dots,\
\chi_n,\ \chi_{-n},\dots \big)$ and
\begin{equation}
  \label{Morb0}
{\cal M}^{\rm KK}\ =\ \left(\! \begin{array}{ccccccc}
0 & m & m & m & m & m & \cdots \\
m & M & 0 & 0 & 0 & 0 & \cdots  \\
m & 0 & M + \frac{1}{R} & 0 & 0 & 0 & \cdots \\
m & 0 & 0 & M - \frac{1}{R} & 0 & 0 & \cdots \\
m & 0 & 0 & 0 & M + \frac{2}{R} & 0 & \cdots \\
m & 0 & 0 & 0 & 0 & M - \frac{2}{R} & \cdots \\
\vdots & \vdots & \vdots & \vdots & \vdots & \vdots & \ddots
\end{array}\!\right)\,,
\end{equation}
with $m = v\bar{h}_1/\sqrt{2}$.  In  a three-generation model, $m$ and
$\bar{h}_1$ are both 3-vectors in  the flavour space, i.e.\ $\bar{h}_1
=  (\bar{h}^e_1,\ \bar{h}^\mu_1,\  \bar{h}^\tau_1)^T$. We will discuss
intergenerational mixing effects in more detail in Section 5. Here, we
assume for simplicity that $\bar{h}_1 = \bar{h}^e_1$.

Following~\cite{DDG2},    we rearrange  the   singlet KK-Weyl  spinors
$\xi_0$ and $\chi^\pm_n$, such that the smallest diagonal entry of the
KK neutrino  mass  matrix  ${\cal   M}^{\rm KK}$ in~(\ref{Morb0})   is
$|\varepsilon|  = {\rm min}\,  \Big(  | M -  \frac{k}{R}  | \Big) \leq
1/(2R)$, for a given value $k=k_0$.   In this newly defined basis, the
effective kinetic Lagrangian~(\ref{Lkin1}) becomes
\begin{equation}
  \label{Lkinorb}
{\cal L}_{\rm kin}\ =\ \frac{1}{2}\, \bar{\Psi}_\nu\,\Big(\, 
i\!\not\!\partial\ -\ {\cal M}^{\rm KK}_\nu\, \Big)\, \Psi_\nu\,,
\end{equation}
where $\Psi_\nu$ is the reordered (4-component) Majorana-spinor vector
\begin{equation}
  \label{Psinu}
\Psi^T_\nu \ =\ \left[\, 
\left(\! \begin{array}{c} \nu_l \\ \bar{\nu}_l \end{array}\!\right)\,,\
\left(\! \begin{array}{c} \chi_{k_0} \\ \bar{\chi}_{k_0} 
                                               \end{array}\!\right)\,,\
\left(\! \begin{array}{c} \chi_{k_0+1} \\ \bar{\chi}_{k_0+1} 
                                               \end{array}\!\right)\,,\
\left(\! \begin{array}{c} \chi_{k_0-1} \\ \bar{\chi}_{k_0-1} 
                                               \end{array}\!\right)\,,\
\cdots\,,
\left(\! \begin{array}{c} \chi_{k_0+n} \\ \bar{\chi}_{k_0+n} 
                                               \end{array}\!\right)\,,\
\left(\! \begin{array}{c} \chi_{k_0-n} \\ \bar{\chi}_{k_0-n} 
                                               \end{array}\!\right)\,,\
\cdots\ \right]
\end{equation}
and ${\cal M}^{\rm KK}_\nu$ the corresponding KK neutrino mass matrix
\begin{equation}
  \label{Morb}
{\cal M}^{\rm KK}_\nu\ =\ \left(\! \begin{array}{ccccccc}
0 & m & m & m & m & m & \cdots \\
m & \varepsilon & 0 & 0 & 0 & 0 & \cdots  \\
m & 0 & \varepsilon + \frac{1}{R} & 0 & 0 & 0 & \cdots \\
m & 0 & 0 & \varepsilon - \frac{1}{R} & 0 & 0 & \cdots \\
m & 0 & 0 & 0 & \varepsilon + \frac{2}{R} & 0 & \cdots \\
m & 0 & 0 & 0 & 0 & \varepsilon - \frac{2}{R} & \cdots \\
\vdots & \vdots & \vdots & \vdots & \vdots & \vdots & \ddots
\end{array}\!\right)\,.
\end{equation}
The  eigenvalues of ${\cal M}^{\rm  KK}_\nu$ can  be computed from the
characteristic eigenvalue  equation  ${\rm   det}\, ( {\cal    M}^{\rm
KK}_\nu - \lambda {\bf 1}) = 0$, which is analytically given by
\begin{equation}
  \label{detorb} 
\prod\limits_{n=0}^\infty\,\bigg[ \Big( \lambda\,
-\,\varepsilon\Big)^2\: -\: \frac{n^2}{R^2}\,\bigg]\,\bigg[\,
1\: +\: \frac{\varepsilon}{\lambda\,
-\,\varepsilon}\ -\ m^2\, \sum\limits_{n=-\infty}^\infty\,\frac{1}{
(\lambda\,-\,\varepsilon)^2\: -\: \frac{n^2}{R^2} }\ \bigg]\ =\ 0\, .
\end{equation}
Since it can be shown that $\lambda - \varepsilon =  \pm n/R$ is never
an  exact  solution to  the  characteristic equation,  only the second
factor in  (\ref{detorb})  can  vanish.   Employing  complex   contour
integration   techniques, the   summation  in the    second factor  in
(\ref{detorb}) can be  performed   exactly, leading to an   equivalent
transcendental equation
\begin{equation}
  \label{transI}
\lambda\ =\ \pi m^2 R\ 
          \cot\, \Big[\,\pi R\, (\lambda - \varepsilon )\,\Big]\, .
\end{equation}
As  was  already discussed  in   \cite{DDG2}, if  $\varepsilon   = 0$,
(\ref{transI})  implies that the mass  spectrum consists of massive KK
Majorana neutrinos degenerate in pairs with opposite  CP parities.  If
$\varepsilon =  1/(2R)$, the  KK  mass  spectrum contains  a  massless
state,  which is predominantly  left-handed   if $mR  < 1$, while  the
remaining massive  KK states  form  degenerate pairs with  opposite CP
parities,  exactly  as in the   $\varepsilon  = 0$ case.   However, if
$\varepsilon   \neq   0,\     1/(2R)$,  the   lepton      number  gets
broken.\footnote{Alternatively, the lepton  number may  also be broken
through   the    Scherk-Schwarz  mechanism, where   the Scherk-Schwarz
rotation  angle will induce  terms   very similar  to those  depending
on~$\varepsilon$~\cite{DDG2,DGQ}.}  In this case, there is no massless
state in   the spectrum,  and the  above   exact degeneracy  among the
massive Majorana   neutrinos  becomes only  approximate, with  a  mass
splitting of   order $2\varepsilon$  for each  would-be  ($\varepsilon
\rightarrow 0$) degenerate KK pair.

We now consider   an  orbifold model,   in  which the $y=0$ brane   is
displaced  from the orbifold  fixed points  by an  amount  $a$.  Under
certain  restrictions in Type I  string theory~\cite{GP,DDG2}, such an
operation can be  performed  respecting the  $Z_2$  invariance  of the
original higher-dimensional action.    In  particular, one   can  take
explicitly account of this last property  by considering the following
replacements in the effective Lagrangian (\ref{Leff}):
\begin{eqnarray}
\xi\, \delta (y - a) &\to & \frac{1}{2}\; \xi\,\Big[\, \delta ( y - a)\: +\:
\delta ( y + a - 2\pi R) \, \Big]\,,\nonumber\\
\eta\, \delta (y - a) &\to & \frac{1}{2}\; \eta\,\Big[\, \delta ( y - a)\: -\:
\delta ( y + a - 2\pi R) \, \Big]\,,
\end{eqnarray}
with $0\le a < \pi R$ and $0\le  y \le 2\pi R$.  It  is obvious that a
$Z_2$-invariant implementation of brane-shifted couplings requires the
existence of two branes at least, placed  at $y= a$ and $y  = 2\pi R -
a$. In addition, we assume that  $a$ is a rational  number in units of
$\pi R$, i.e.
\begin{equation}
  \label{rat}
 a\ =\ \frac{r}{q}\ \pi R\; ,
\end{equation}
where   $r,q$  are natural numbers.   This  last  assumption  has been
introduced  for  technical reasons.     It enables  us to  carry   out
analytically the infinite summations  over  KK  states (see also   our
discussion below).

Proceeding  as  above, the effective   KK neutrino mass  matrix ${\cal
M}^{\rm KK}_\nu$ for  the orbifold model  with a shifted brane can  be
written down in an analogous form
\begin{equation}
  \label{Morbshift}
{\cal M}^{\rm KK}_\nu\ =\ \left(\! \begin{array}{ccccccc}
0 & m^{(0)} & m^{(1)} & m^{(-1)} & m^{(2)} & m^{(-2)} & \cdots \\
m^{(0)} & \varepsilon & 0 & 0 & 0 & 0 & \cdots  \\
m^{(1)} & 0 & \varepsilon + \frac{1}{R} & 0 & 0 & 0 & \cdots \\
m^{(-1)} & 0 & 0 & \varepsilon - \frac{1}{R} & 0 & 0 & \cdots \\
m^{(2)} & 0 & 0 & 0 & \varepsilon + \frac{2}{R} & 0 & \cdots \\
m^{(-2)} & 0 & 0 & 0 & 0 & \varepsilon - \frac{2}{R} & \cdots \\
\vdots & \vdots & \vdots & \vdots & \vdots & \vdots & \ddots
\end{array}\!\right)\,,
\end{equation}
where 
\begin{eqnarray}
  \label{mk0}
m^{(n)} &=& \frac{v}{\sqrt{2}}\, \bigg[\, \bar{h}_1\, 
\cos\bigg(\frac{ (n-k_0) a}{R}\bigg)\: +\: \bar{h}_2\, 
\sin\bigg(\frac{ (n-k_0) a}{R}\bigg)\,\bigg]\ =\
m\,\cos\bigg(\frac{na}{R}\, -\, \phi_h\,\bigg)\,,\qquad
\end{eqnarray}
with $m  =  v  \sqrt{(\bar{h}^2_1  + \bar{h}^2_2)/2}$   and  $\phi_h =
\tan^{-1} (\bar{h}_2 / \bar{h}_1) + k_0 a /R$.  As before, we consider
an one-generation model  with $\bar{h}_1 = \bar{h}^e_1$ and $\bar{h}_2
= \bar{h}^e_2$,  which    renders the analytic   determination of  the
eigenvalue   equation  tractable.  We will  relax   this assumption in
Section 5,  when   discussing  the compatibility of   this  model with
neutrino oscillation data.  Thus, for our one-generation brane-shifted
model, the characteristic eigenvalue equation reads
\begin{equation}
  \label{detorbshift} 
\prod\limits_{n=0}^\infty\,\bigg[ \Big( \lambda\,
-\,\varepsilon\Big)^2\: -\: \frac{n^2}{R^2}\,\bigg]\,\bigg[\,
1\: +\: \frac{\varepsilon}{\lambda\,
-\,\varepsilon}\ -\ \frac{1}{\lambda - \varepsilon}\,
\sum\limits_{n=-\infty}^\infty\,\frac{m^{(n)\, 2}}{
\lambda\: -\: \varepsilon\: -\: \frac{n}{R}\:   }\ \bigg]\ =\ 0\, ,
\end{equation}
which is equivalent to
\begin{equation}
  \label{eigen}
\lambda\ =\ \sum\limits_{n=-\infty}^\infty\;
\frac{m^{(n)\, 2}}{\lambda\: -\: \varepsilon\: -\: \frac{n}{R}}\ .
\end{equation}
As opposed to the  $a=0$ case, complex contour  integration techniques
are  not directly     applicable  in  evaluating    the  infinite  sum
in~(\ref{eigen}).    The  preventive reason   is  that   the  function
$m^{(n)}$, analytically continued to  the  complex $n$-plane, is   not
bounded from above as $n \to \pm i\infty$, as it had to be, because of
its dependence on $\cos (na/R)$.  However, as has been mentioned above
and  discussed   further in    Appendix  A, this  difficulty   may  be
circumvented by assuming that $a$  is a rational   number in units  of
$\pi R$, as stated  in~(\ref{rat}).  Under this  technical assumption,
we    carry out   in  Appendix~A  the   infinite sum  in~(\ref{eigen})
analytically and derive the eigenvalue equation for the simplest class
of cases,  where $a = \pi  R/q$ with $r=1$  and $q$  an integer larger
than 1,    i.e.\  $q\ge 2$.    More precisely,   we find~\footnote{The
so-derived formula  generalizes the  one  presented  in~\cite{DDG2} to
include brane-shifting and arbitrary Yukawa-coupling effects.}
\begin{eqnarray}
  \label{transII}
\lambda &=& \pi m^2 R\ \bigg\{ \cos^2 \Big[\, \phi_h\, -\, a (\lambda
- \varepsilon) \,\Big]\, \cot\Big[\,\pi R\, (\lambda - \varepsilon )\,\Big]
\ -\  \frac{1}{2}\sin \Big[\, 2\phi_h\, -\, 2a (\lambda
- \varepsilon) \,\Big]\ \bigg\}\, .\qquad
\end{eqnarray}
Observe that unless $\varepsilon = 1/(2R)$, $a =  \pi R/2$ and $\phi_h
= \pi/4$,   the mass spectrum  consists   of massive non-degenerate KK
neutrinos.   However, it can   be shown from~(\ref{transII}) that this
tree-level mass  splitting between a pair of  KK Majorana neutrinos is
generally small for $m_{(n)} \gg 1/R$.  In particular, this tree-level
mass splitting is almost  independent of $a$  and subleading so  as to
play any relevant r\^ole in our calculations.

At this stage, it is important to comment on taking the limit $a = \pi
R/ q \to  0$ in~(\ref{transII}), or equivalently $q  \to \infty$. This
limit is not the eigenvalue equation~(\ref{transI}) which is valid for
$a=0$, because  of the presence  of the extra non-vanishing  term that
depends on  $\sin(2\phi_h)$ in~(\ref{transII}). This  apparent paradox
can  be resolved  by  noticing  that the  existence  of this  would-be
anomalous  term is  ensured only  if  the brane-shifting  $a$ is  much
larger than  the fundamental quantum gravity scale  $M_F$, i.e.\ $a\gg
1/M_F$. Since  $M_F$ represents a natural ultra-violet  cut-off of the
theory, we expect the onset of new physics above the scale $M_F$, most
likely of stringy nature, effectively implying that the KK-Yukawa mass
terms $m^{(n)}$ are exponentially suppressed or zero for KK-numbers $n
\stackrel{>}{{}_\sim} M_F  R$.  As  we will explicitly  demonstrate in
Section 4 (see our discussion in (\ref{mnueff})), such a truncation of
the  KK sum  at $M_F$  effectively results  in a  modification  of the
eigenvalue equation~(\ref{transII}) to
\begin{eqnarray}
  \label{transIII}
\lambda &=& m^2 R\, \bigg\{ \pi\, \cos^2 \Big[\, \phi_h\, -\, a (\lambda
- \varepsilon) \,\Big]\, \cot\Big[\,\pi R\, (\lambda - \varepsilon
)\,\Big]\nonumber\\
&& -\ {\rm Si} (2aM_F)\, \sin \Big[\, 2\phi_h\, -\, 2a (\lambda
- \varepsilon) \,\Big]\ \bigg\}\, .\qquad
\end{eqnarray}
In the  above, ${\rm Si} (x)  = \int_0^x dt\,\frac{\sin  t}{t}$ is the
integral-sine function.  For any finite  value of its  argument, ${\rm
Si}(x)$ can be expanded as
\begin{equation}
  \label{Si} 
{\rm Si} (x)\  =\ \sum_{n=1}^{+\infty}\,
\frac{(-1)^{(n-1)}\,x^{(2n -1)}}{(2n-1)\,(2n-1)!}\ .
\end{equation}
For small $x$, it  is ${\rm Si} (x) \approx x$, while  ${\rm Si} (x) =
\pi/2$ for  $x \to \infty$.   Clearly, as long  as $a \gg  1/M_F$, the
eigenvalue  equations~(\ref{transII}) and (\ref{transIII})  are almost
identical,   since  ${\rm  Si}(2aM_F)   =  \pi/2$   to  a   very  good
approximation. On the other hand, the limit $a\to 0$ does now smoothly
go over to~(\ref{transI}), as it should be.

Finally, in addition to  the aforementioned tree-level mass splitting,
one-loop radiative effects may also contribute to further increase the
mass difference  between two nearly degenerate  KK Majorana neutrinos,
if  $\bar{h}_1$ and  $\bar{h}_2$  do not  vanish simultaneously.   The
one-loop  generated  mass  splitting,   however,  is  expected  to  be
small~\cite{AP1}  of  order  $\bar{h}_1\bar{h}_2m_{(n)}/(8\pi^2)  \sim
10^{-2}   \times  (M_F/M_P)^2  \times   m_{(n)}  \stackrel{<}{{}_\sim}
10^{-2}\times \Delta m_{(n)}$, where $m_{(n)} \approx n/R \leq M_F$ is
the  approximate  mass of  the  $n$th  KK  pair of  nearly  degenerate
Majorana neutrinos, and $\Delta  m_{(n)} = m_{(n+1)} - m_{(n)} \approx
1/R$ is  the mass difference  between two adjacent KK  Majorana pairs.
Although  such  a  radiatively-induced   mass  splitting  may  play  a
significant  r\^ole  for leptogenesis~\cite{AP1},  its  effect on  the
double beta  decay  amplitude is  negligible.   Therefore, we  neglect
radiative effects on the KK mass spectrum throughout the paper.

\setcounter{equation}{0}
\section{RG evolution of neutrino Yukawa couplings}

The RG evolution of the Yukawa couplings in the standard 4-dimensional
scenario  involving    sterile  neutrinos   has    been discussed   in
\cite{pokorski}.  Here,  we derive the  corresponding RG equations for
the  higher-dimensional case.   Since the  RG  evolution equations for
$\bar{h}^l_1$ and $\bar{h}^l_2$ will  be similar, we  concentrate only
on the former ($\equiv \h$).   In such a higher-dimensional  scenario,
the  presence of the KK   sterile states alters   the RG running.  The
triangle and self-energy  diagrams   that contribute to   the  running
remain the  same as in the SM,  except that in the  higher dimensional
context,  wherever there are  internal   $\xi_n$  lines,  there is   a
multiplicative  factor  $t_\delta =   (\mu  R)^\delta X_\delta$,  with
$X_\delta = 2\pi^{\delta/2}/\delta\Gamma(\delta/2)$.  The RG  equation
for the Yukawa coupling $\h$ is given by
\begin{eqnarray}
\label{diffeqn} 
16\pi^2 \frac{d\h}{d\ln\mu} & = & \frac{3}{2}\, \left[\, t_\delta\,(\h
\h^\dagger)\, \h - \h\, (h_e^\dagger h_e)\, \right]\: +\: \h\, {\rm
Tr} \left(\, 3 h_u^\dagger h_u + 3 h_d^\dagger h_d + h_e^\dagger h_e +
t_\delta\, \h^\dagger \h \right)\nonumber\\ &&-\: \h\, \left(\,
\frac{9}{4} g_w^2 + \frac{3}{4} g'^2\, \right),
\end{eqnarray}
where  $g_w$  and $g'$ are  the  SU(2)$_L$ and U(1)$_Y$ gauge-coupling
constants,  respectively.  Note  that for   $\delta  =  0~(1)$, it  is
$X_\delta  = 1~(2)$.   Also,  for $\delta  = 0$,  $t_\delta = 1$,  the
standard RG equation is reproduced \cite{pokorski}.

We  now observe that   the four-dimensional Yukawa  coupling ($\h$) is
suppressed  with respect to  the  higher-dimensional coupling ($h$) by
means of the relation: $\h =  (M_F/M_P)^{\delta/n_g} h$. Thus, even if
we consider $h (1/R) \sim  1$, the four-dimensional $\h$ is suppressed
by many  orders of magnitude.   {}From Eq.~(\ref{diffeqn}), it is also
obvious that  unless $t_\delta$ is large enough  to be comparable with
$(M_P^2/M_F^2)^{\delta/n_g}$,     the contributions from the top-quark
Yukawa coupling or the gauge couplings dominate the running, and hence
there is no power-law behaviour at lower energies.

On the contrary,  if we  go  to a very high  energy  such that  we can
ignore $h_t$, then the terms multiplying  $t_\delta$ dominate. In such
a case, ignoring the gauge contribution, we can write
\begin{eqnarray}
\label{approxeqn}
16\pi^2 \frac{d\h}{d\ln\mu}\ \sim\ \frac{5}{2}\; t_\delta\, \h^3\, .  
\end{eqnarray}
Integrating Eq.~(\ref{approxeqn}) from the scale $\mu_0 \equiv R^{-1}$
to $\mu$, we obtain
\begin{eqnarray}
\label{eqn2}
\frac{1}{\bar{h}^2(1/R)}\ -\ \frac{1}{\bar{h}^2(\mu)}\ \simeq\ \frac{5
X_\delta}{16\pi^2\delta}\; (\mu R)^\delta\, .  
\end{eqnarray}
In  terms  of the  Yukawa  fine  structure  constant $\alpha  (\mu)  =
h^2(\mu)/(4\pi)$ of  the original 5-dimensional  Yukawa coupling ($h$)
and for the simple case $\delta = n_g$, (\ref{eqn2})~takes on the form
\begin{eqnarray}
  \label{master}
\frac{1}{\alpha (\mu)}\ \simeq\ 
\frac{1}{\alpha (1/R)}\ -\ \frac{5
X_\delta}{4\pi \delta}\, \left(\frac{\mu}{M_F}\right)^\delta \;. 
\end{eqnarray} 
Clearly, $\alpha (\mu) \rightarrow \infty$, for a critical scale 
\begin{eqnarray}
  \label{landau}
\mu_{\rm critical}\ =\ M_F\, \left(\frac{4\pi \delta}{5X_\delta\, 
\alpha (1/R)}\right)^{1/\delta} 
\end{eqnarray}
Interestingly  enough,  (\ref{landau})   implies  that  the  power-law
behaviour sets in not  just above the compactification scale $R^{-1}$,
as  was  naively  expected~\cite{DDG2},  but well  above  the  quantum
gravity scale $M_F$.  On the  other hand, requiring that $\alpha (M_F)
\le  1$ in  (\ref{master})  implies  that $\alpha  (1/R)  < 0.55$  for
$\delta  =1$.  This  last condition  assures that  our  theory remains
perturbative  up  to the  quantum  gravity  scale  $M_F$.  {}From  our
discussion above, it  is obvious that power-law effects  on the Yukawa
neutrino couplings can be safely neglected in our analysis.

\setcounter{equation}{0}
\section{Effective neutrino-mass estimates}

In this section, we calculate the $0\nu\beta\beta$ observable $\langle
m\rangle$ in orbifold  5-dimensional models.  This quantity determines
the size  of the neutrinoless  double  beta decay amplitude,  which is
induced by $W$-boson exchange graphs.  To this end, it is important to
know the  interactions of the  $W^\pm$  bosons to the  charged leptons
$l=e, \mu,  \tau$  and  the KK-neutrino   mass-eigenstates  $n_{(n)}$.
Adopting the  conventions of~\cite{IP1}, the effective charged current
Lagrangian is given by
\begin{equation}
  \label{charged}
{\cal L}^{W^\pm}_{\rm int}\ =\ - \frac{g_w}{\sqrt{2}}\,
W^{-\mu}\, \sum_{l=e,\mu ,\tau}\, \bigg( B_{l\nu_l}\,
\bar{l}\,\gamma_\mu P_L\, \nu_l\: +\: \sum_{n=-\infty}^{+\infty}\,
B_{l,n}\, \bar{l}\,\gamma_\mu P_L\, n_{(n)}\,\bigg)\: +\: {\rm h.c.}\,,
\end{equation}
where $g_w$ is the weak  coupling constant, $P_L = (1-\gamma_5)/2$  is
the  left-handed chirality   projector,    and $B$  is    an  infinite
dimensional mixing matrix.  The   matrix $B$ satisfies  the  following
crucial identities:
\begin{eqnarray}
  \label{B1}
B_{l\nu_l} B^*_{l'\nu_l}\: +\: \sum_{n=-\infty}^{+\infty}\,
B_{l,n} B^*_{l',n} &=& \delta_{ll'}\,,\\
  \label{B2}
B_{l\nu_l}\, m_{\nu_l}\, B_{l'\nu_l}\: +\: \sum_{n=-\infty}^{+\infty}\,
B_{l,n}\, m_{(n)}\, B_{l',n} &=& 0\,.
\end{eqnarray}
Equation (\ref{B1}) reflects  the unitarity properties of  the charged
lepton weak  space,  and (\ref{B2})  holds true, as   a result  of the
absence of the Majorana mass terms $\nu_l \nu_{l'}$ from the effective
Lagrangian in the flavour basis.  For the models under discussion, the
KK neutrino masses $m_{(n)}$     can  be determined exactly  by    the
solutions  of the corresponding   transcendental equations.  To a good
approximation,  however,  these  solutions   for large   $n$  simplify
to\footnote{For   $|n| >\varepsilon$ and     $n  < 0$,  the  KK   mass
eigenvalues $m_{(n)}$ are  negative.   This corresponds to a  neutrino
with positive physical mass  $|m_{(n)}|$ and negative  CP parity.  One
can  always take account  of the negative CP  parity by redefining the
mixing matrix elements  $B_{l,-n}$ as $B_{l,-n}  \to i  B_{l,-n}$, for
$n>\varepsilon R >0$. Although  we will allow negative neutrino masses
in our calculations, we  should stress that  both approaches are fully
equivalent leading to the same analytic results.}
\begin{equation}
  \label{mn}
m_{(n)}\ \approx\  \frac{n}{R}\ + \: \varepsilon\; .
\end{equation}
This last    expression proves to    be a  good  approximation in  our
estimates.

According to~(\ref{t1/2}), the $0\nu\beta\beta$-decay amplitude ${\cal
T}_{0\nu\beta\beta}$ is given by~\cite{Klapdor}:
\begin{equation}
  \label{M2beta}
{\cal T}_{0\nu\beta\beta}\ =\ \frac{\langle m \rangle}{m_e}\ 
                         {\cal M}_{\rm GTF} (m_\nu )\; ,
\end{equation}
where ${\cal M}_{\rm  GTF} = {\cal M}_{\rm GT}  - {\cal M}_{\rm F}$ is
the  difference of  the   nuclear matrix  elements for   the so-called
Gamow-Teller  and  Fermi transitions.   Note   that this difference of
nuclear  matrix  elements  sensitively  depends   on the  mass  of the
exchanged KK neutrino in  a $0\nu\beta\beta$ decay. Especially if  the
exchanged KK-neutrino mass $m_{(n)}$ is comparable  or larger than the
characteristic  Fermi  nuclear   momentum $q_F  \approx 100$~MeV,  the
nuclear   matrix element    ${\cal     M}_{\rm  GTF}$  decreases    as
$1/m^2_{(n)}$.    The  general     expression  for    the    effective
Majorana-neutrino mass $\langle  m \rangle$ in~(\ref{M2beta}) is given
by
\begin{equation}
  \label{effMajmass}
\langle m \rangle\ =\ \frac{1}{{\cal M}_{\rm GTF} (m_\nu) }\
\sum_{n = -\infty}^{\infty}\, B^2_{e,n}\,m_{(n)}\, 
\Big[\, {\cal M}_{\rm GTF} (m_{(n)})\: 
-\: {\cal M}_{\rm GTF} (m_\nu )\, \Big]\; .\quad
\end{equation}
In the above, the  first term describes the genuine higher-dimensional
effect of KK-neutrino exchanges, while the second term is the standard
contribution  of  the  light   neutrino  $\nu$,  rewritten  by  virtue
of~(\ref{B2}).  Note that the dependence of the nuclear matrix element
${\cal  M}_{\rm GTF}$  on the  KK-neutrino masses  $m_{(n)}$  has been
allocated to  $\langle m  \rangle$ in (\ref{effMajmass}).   The latter
generally leads to predictions for  $\langle m \rangle$ that depend on
the  double beta  emitter isotope  used in  experiment.   However, the
difference in the predictions  is too small for the higher-dimensional
singlet-neutrino models  to be  able to operate  as a smoking  gun for
different $0\nu\beta\beta$-decay experiments.

\subsection{Factorization Ansatz for analytic estimates}

To obtain analytic  estimates  that will help  us   to gain  a  better
insight into the dynamical properties of~(\ref{effMajmass}), it proves
useful   to  approximate the $0\nu\beta\beta$-decay  amplitude  ${\cal
T}_{0\nu\beta\beta}$ in~(\ref{M2beta})  by means  of the  factorizable
Ansatz~\cite{HK}:
\begin{equation}
  \label{FA}
{\cal T}_{0\nu\beta\beta}\ \approx\ 
\frac{\langle m \rangle_{\rm SA}}{m_e}\ {\cal M}_{\rm GTF} 
(m_\nu )\ +\ \frac{m^2_p}{m_e}\, 
\langle m^{-1} \rangle\, {\cal M}_{\rm GTF} (m_p)\, ,
\end{equation}
where $m_p$ is the proton mass, and ${\cal  M}_{\rm GTF} (m_\nu )$ and
${\cal M}_{\rm GTF}  (m_p)$  are the   values  of the nuclear   matrix
element ${\cal  M}_{\rm   GTF}$ at $m_\nu$  and   $m_p$, respectively.
In~(\ref{FA}), the $0\nu\beta\beta$ matrix element has been written as
a sum  of  two terms. The  first   term, which  is the  dominant  one,
accounts   for  effects coming from    KK  neutrinos lighter than  the
characteristic Fermi nuclear momentum $q_F  \approx 100$~MeV. In  this
kinematic region, the nuclear  matrix element ${\cal M}_{\rm GTF}$  is
almost independent of the KK neutrino mass  $m_{(n)}$. The second term
in~(\ref{FA}) is due to KK neutrinos much heavier than $q_F$.  This is
generically a subdominant contribution to ${\cal T}_{0\nu\beta\beta}$,
since ${\cal M}_{\rm GTF} (m_p) \ll {\cal M}_{\rm GTF} (m_\nu)$.

The quantity  $\langle m \rangle_{\rm SA}$ is  an approximation of the
effective Majorana-neutrino  mass   $\langle   m \rangle$, which    is
obtained by approximating the  nuclear matrix elements  ${\cal M}_{\rm
GTF} (m_{(n)})$ entering $\langle m  \rangle$ in (\ref{effMajmass}) by
a step function at $|m_{(n)}| = q_F$:
\begin{equation}
  \label{SA}
{\cal M}_{\rm  GTF} (m_{(n)})\ =\ \left\{
\begin{array}{r}
{\cal  M}_{\rm GTF} (m_\nu  )\,, \qquad \mbox{for}\ |m_{(n)}| \le q_F\;,\\
0\,,\qquad \mbox{for}\ |m_{(n)}| > q_F\;. \end{array} \right.
\end{equation}
In  what follows, we refer  to such an approach  to the nuclear matrix
elements as the  Step Approximation~(SA).  The effective neutrino mass
in the SA reads:
\begin{eqnarray}
  \label{effmass}
\langle m \rangle_{\rm SA} & = & B^2_{e\nu} m_\nu\ + \sum_{n =
-[(q_F+\varepsilon)R]}^{[(q_F-\varepsilon)R]} B^2_{e,n}\,m_{(n)}\nonumber\\
& = & -\, \sum_{n = [(q_F-\varepsilon)R]}^{+\infty}\, B^2_{e,n}
m_{(n)}\ - \sum_{n = [(q_F+\varepsilon)R]}^{+\infty}\, 
B^2_{e,-n} m_{(-n)}\; ,
\end{eqnarray}
where  we used  (\ref{B2})  to arrive  at  the last  equality for  the
effective neutrino mass. Notice that  $\langle m \rangle$ is not zero,
simply because  the sum  over the KK  neutrino states is  truncated to
those with a mass $|m_{(n)}|, |m_{(-n)}| \le q_F$. 

Correspondingly, the effects of  the heavier KK neutrinos, with masses
$m_{(n)} \stackrel{>}{{}_\sim}  q_F$, have been taken  into account in
the factorizable  Ansatz~(\ref{FA}) by means of  the inverse effective
neutrino  mass  $\langle   m^{-1}  \rangle$.   This  newly  introduced
quantity is given by
\begin{equation}
  \label{effmass2}
\langle m^{-1} \rangle\ =\ 
\sum_{n = [(q_F-\varepsilon) R]}^{+\infty}\, B^2_{e,n}
m^{-1}_{(n)}\  + \sum_{n = [(q_F+\varepsilon) R]}^{+\infty}\, 
B^2_{e,-n} m^{-1}_{(-n)}\;  .
\end{equation}
The factorizable form~(\ref{M2beta}) of the matrix element constitutes
a  good approximation except for the  isolated region where $|m_{(n)}|
\approx   q_F \approx 100$~MeV. Nevertheless,    the effect of the  KK
neutrinos on the effective neutrino  mass is cumulative~\cite{IP}  due
to  a sum of  an infinite number  of states,  since  each KK state has
either a  tiny Majorana  mass or a   very suppressed  mixing  with the
electron neutrinos. Therefore, we  expect that excluding this isolated
region of KK-neutrino   contributions  around  $q_F$ will  not   alter
quantitatively our results in a relevant way.

We will now rely on (\ref{effmass}) to estimate the effective neutrino
mass   $\langle  m   \rangle_{\rm  SA}$   in  different   settings  of
5-dimensional orbifold models discussed  in Section~2.  To begin with,
let us consider a simple orbifold model, with $\varepsilon \neq 0$ and
$\varepsilon \neq 1/(2R)$.  In addition, we consider the case $a = 0$,
namely we take the brane to be  located at the one of the two orbifold
fixed points.   Like the  neutrino masses, the  mixing-matrix elements
$B_{e\nu}$ and $B_{e,n}$ can also be computed exactly \cite{DDG2}:
\begin{eqnarray}
  \label{Bln} 
B_{e\nu} & = &   \frac{1}{\cal N}\,,\qquad B_{e,n} \  =\
  \frac{1}{{\cal N}_{(n)}}\ ,
\end{eqnarray}
where the squares of the  normalization factors ${\cal N}$ and ${{\cal
N}_{(n)}}$ are given by
\begin{eqnarray}
  \label{Norm}
{\cal N}^2\ =\ 1\: +\: \sum_{n=-\infty}^{+\infty}
\frac{m^2}{\Big(\,\varepsilon\, -\, m_\nu\, +\, \frac{n}{R}\,\Big)^2}\  
,\qquad
{{\cal N}^2_{(n)}}\ =\ 1\: +\: \sum_{k=-\infty}^{+\infty}
\frac{m^2}{\Big(\,\varepsilon\, -\, m_{(n)}\, +\, \frac{k}{R}\,\Big)^2}\ . 
\end{eqnarray}
Applying complex integration methods for convergent infinite sums, the
squared normalization factor ${\cal N}^2$ can be calculated to give
\begin{equation}
  \label{Nexact} {\cal N}^2\ =\ 1\: +\: \frac{\pi^2 m^2 R^2}{\sin^2[\pi
R(m_\nu -\varepsilon)]}\ =\ 1\: +\: \pi^2 m^2 R^2\: +\:
\frac{m^2_\nu}{m^2}\ .
\end{equation}
In obtaining  the last   equality  in  (\ref{Nexact}),  we   used  the
eigenvalue  equation  (\ref{transI}) for  $\lambda  =  m_\nu$.  {}From
(\ref{Bln}) and (\ref{Nexact}), we immediately see that  if $mR \ll 1$
and $m_\nu \ll m$, it is  $B_{e\nu} \approx 1$  and hence the lightest
neutrino state is  predominantly left-handed.  For the calculation  of
the effective neutrino mass, we need
\begin{eqnarray} 
\label{Nnexact} 
{{\cal N}^2_{(n)}} \ =\ 1\: +\: \frac{\pi^2 m^2 R^2}{\sin^2[\pi
R(m_{(n)} -\varepsilon)]}\ =\ 1\: +\: \pi^2 m^2 R^2\: +\:
\frac{m^2_{(n)}}{m^2}\ \approx\ 
\frac{(\,\frac{n}{R} + \varepsilon\,)^2}{m^2}\ ,
\end{eqnarray}
where the last  approximate equality in (\ref{Nnexact}) corresponds to
a large  $n$.   In Appendix~B, we   show that the KK   neutrino masses
derived  from~(\ref{transI})  and   the  mixing-matrix elements  given
in~(\ref{Bln})  satisfy        the sum     rules     given    by   the
identities~(\ref{B1}) and (\ref{B2}).

Based  on~(\ref{effmass}), we  will  now  perform  an estimate  of the
effective neutrino mass in  the simple orbifold model mentioned above.
Plugging   the      value   of    $B_{e,n}    =     1/{\cal  N}_{(n)}$
into~(\ref{effmass}), we may  estimate the effective neutrino  mass in
the SA through the following steps:
\begin{eqnarray}
  \label{mnu1}
\langle m \rangle_{\rm SA} &= & -m^2\, \sum_{n = [q_F R]}^\infty\, \bigg(\,
\frac{1}{\varepsilon\, +\, \frac{n}{R} }\ + \ \frac{1}{\varepsilon \,
-\, \frac{n}{R}}\ \bigg)\ +\ 
{\cal O}\Bigg(\frac{\varepsilon m^2 R}{q_F}\Bigg) \nonumber\\
&\approx&  m^2R\, \int\limits_{q_FR}^{+\infty} dn\,  \bigg(\,
\frac{1}{n\, -\, \varepsilon R }\ - \ \frac{1}{n \,
+\, \varepsilon R}\ \bigg)\ =\ -\, m^2 R\ 
\ln\bigg(\frac{q_F-\varepsilon}{q_F+\varepsilon}\bigg)\nonumber\\
&= & {\cal O}\Bigg(\frac{\varepsilon m^2 R}{q_F}\Bigg)\ . 
\end{eqnarray}
In arriving at the last  equality in (\ref{mnu1}), we approximated the
sum  over  the KK  states  by  an integral,  and  used  the fact  that
$\varepsilon/    q_F    \ll    1$.     Since    $2    \varepsilon    R
\stackrel{<}{{}_\sim}  1$,  we can  estimate  that  for  $m =  10$~eV,
$\langle m  \rangle_{\rm SA} \stackrel{<}{{}_\sim}  10^{-6}$~eV, which
is undetectably small.

The above large suppression of  the effective neutrino mass $\langle m
\rangle_{\rm SA}$  is a consequence of the  very drastic cancellations
due to KK  neutrinos with opposite CP-parities.  However,  we might be
able to  overcome this difficulty by arranging  the opposite CP-parity
KK  neutrinos to couple  to the  electron and  $W$ boson  with unequal
strength.   In   fact,  this  is  what  happens   in  orbifold  models
automatically, if the $y=  0$ brane is shifted to $y =  a \neq 0$.  In
this  case, the  mixing-matrix elements  $B_{e\nu}$ and  $B_{e,n}$ are
given by the inverse of  ${\cal N}$ and ${\cal N}_{(n)}$ respectively;
but now for the shifted brane, ${\cal N}_{(n)}$ is given by
\begin{eqnarray}
  \label{Nnfinal}
{{\cal N}^2_{(n)}}\ =\ 1\: +\: m^2 \sum_{k=-\infty}^{+\infty}
\frac{\cos^2\Big(\frac{ka}{R} - \phi_h\Big)}
{\Big(\,\varepsilon\, -\, m_{(n)}\, +\, \frac{k}{R}\,\Big)^2} \
\approx\ \frac{(\,\frac{n}{R} + \varepsilon\,)^2}
{m^2 \cos^2(\, \frac{na}{R} - \phi_h\,)}\ ,
\end{eqnarray}
where the  second approximate  equality in (\ref{Nnfinal}) corresponds
to large $n$.

By   analogy   to~(\ref{mnu1}),   we  may     compute the    effective
Majorana-neutrino mass for the the brane-shifted scenario ($a \neq 0$)
as follows:
\begin{eqnarray}
  \label{mnu2}   
\langle  m \rangle_{\rm SA}   &\approx  &      
- \, m^2 R\, \int\limits_{q_FR}^{+\infty}  dn\,            \bigg(\,
\frac{\cos^2\Big(\frac{na}{R} - \phi_h\Big)   }{n\,
+\,   \varepsilon   R    }\  -    \   \frac{\cos^2\Big(\frac{na}{R} 
+ \phi_h\Big) }{n \, -\,  \varepsilon  R}\ \bigg)\nonumber\\  
&=& -\, \sin (2\phi_h)\,  m^2 R\,
\int\limits^{M_F R}_{q_F R}  \frac{dn}{n}\
\sin\bigg(\frac{2na}{R}\bigg)\quad   +\quad 
{\cal O}\Bigg(\frac{\varepsilon m^2 R}{q_F}\Bigg)\, .
\end{eqnarray}
In the second step, we have truncated the  upper limit of the integral
at  the   fundamental quantum gravity scale   $M_F$.   The scale $M_F$
represents a  natural ultra-violet cut-off of  the  problem, beyond of
which the onset of  string-threshold  effects are expected to   occur.
The last result in~(\ref{mnu2}) can  now be expressed  in terms of the
integral-sine function ${\rm Si} (x) = \int_0^x dt\,\frac{\sin t}{t}$.
Thus, the effective neutrino mass can be given by
\begin{equation}
  \label{mnueff}
\langle  m \rangle_{\rm SA}\ \approx\ -\, \sin(2\phi_h) \,
m^2 R\ \bigg[\, {\rm Si} (2aM_F)\ -\ {\rm Si} (2aq_F)\, \bigg]\quad 
+ \quad {\cal O}\Bigg(\frac{\varepsilon
m^2 R}{q_F}\Bigg)\, .
\end{equation}
Notice  that  for   a  fixed  given  value  of   $M_F$,  the  analytic
expression~(\ref{mnueff})  for   the  effective  neutrino   mass  goes
smoothly to~(\ref{mnu1}) in  the limit $a\to 0$, as  it should be.  In
order that  the prediction for neutrinoless  double beta decay effects
is at the level reported  recently~\cite{HMexp}, we only need to have:
$\phi_h  \sim  \pm \pi  /4$  and  $1/M_F  \ll a  \stackrel{<}{{}_\sim}
1/(2q_F)$, i.e.\ the brane is slightly displaced from its origin.  For
instance, if $a \approx 1/(3q_F)$, $m  = 10$~eV and $1/R = 300$~eV, we
find that  $\langle m \rangle_{\rm  SA}$ is exactly at  the observable
level, i.e.\ $\langle m \rangle_{\rm SA} \sim 0.4$~eV.

It is  now interesting to  give  an estimate of the  inverse effective
neutrino mass $\langle m^{-1} \rangle$   in the orbifold model with  a
shifted brane ($a\neq 0$).  The quantity  $\langle m^{-1} \rangle$ can
be approximately calculated as follows:
\begin{eqnarray}
  \label{mninverse} 
\langle   m^{-1}   \rangle \!\!&\approx &\!\!    m^2 R^3\,
\int\limits_{q_FR}^{+\infty}               dn\,               \bigg(\,
\frac{\cos^2\Big(\frac{na}{R}   - \phi_h\Big)  }{(n\, +\,  \varepsilon
R)^3 }\ - \ \frac{\cos^2\Big(\frac{na}{R} + \phi_h\Big) }{(n
\,   -\, \varepsilon  R)^3}\   \bigg)\\ 
&&\hspace{-1cm}=\  \sin(2\phi_h)\; m^2  R^3\,
\int\limits_{q_FR}^{+\infty}                           \frac{dn}{n^3}\
\sin\bigg(\frac{2na}{R}\bigg)\:  +\:   \frac{3}{2}\cos(2\phi_h)\;   m^2
\varepsilon   R^4\,      \int\limits_{q_FR}^{+\infty}  \frac{dn}{n^4}\
\sin\bigg(\frac{2na}{R}\bigg)\: -\: \frac{\varepsilon m^2  R}{2q^3_F}\
.  \nonumber
\end{eqnarray}
The RHS of the last equality  in~(\ref{mninverse}) can be written down
in a lengthy expression in terms of the integral-sine, integral-cosine
and known trigonometric functions.  For example, for $\phi_h = \pi/4$,
$\langle m^{-1} \rangle$ is given by
\begin{eqnarray}
  \label{minv2}
\langle  m^{-1}\rangle   \!\!&\approx&\!\!  
2m^2R\,\bigg[\, a^2\,\bigg( {\rm Si}\, (2aq_F)\: -\:
\frac{\pi}{2}\,\bigg)\ -\ \frac{1}{4q^2_F}\, \sin (2aq_F)\ -\ \frac{a}{2q_F}\,
\cos(2aq_F)\, \bigg]\ -\ \frac{\varepsilon m^2 R}{2q^3_F}\; .\qquad \nonumber\\
&&
\end{eqnarray}
For the   specific model considered above,  with  $m = 10$~eV,  $1/R =
300$~eV and $a   =  1/(3q_F)$, we   find that $\langle   m^{-1}\rangle
\stackrel{<}{{}_\sim} 10^{-5}$~TeV$^{-1}$.   Hence, the above exercise
shows that the     contribution from $\langle m^{-1}\rangle$   to  the
double beta decay   amplitude~(\ref{M2beta})  is subdominant;  it gets
even more suppressed for $a \ll 1/q_F$.

\subsection{Numerical evaluation}

To obtain  realistic predictions for the  double beta decay observable
$\langle m  \rangle$, one has to  take into account the  dependence of
${\cal M}_{\rm GTF}$ on the KK neutrino masses $m_{(n)}$.  To properly
implement   this  $m_{(n)}$-dependence  in    our extractions of   the
effective Majorana mass $\langle m \rangle$ from the different nuclei,
we  have  used   the  general  formula~(\ref{effMajmass}),   where the
infinite sum  over $n$ has been truncated  at $|n_{\rm max}| = M_F R$,
namely at  the quantum gravity scale  $M_F$.  Notice  that the general
formula  for  $\langle m  \rangle$  in~(\ref{effMajmass}) includes the
contributions  from the KK neutrinos heavier  than $q_F$, described by
the inverse  effective     neutrino mass    $\langle    m^{-1}\rangle$
in~(\ref{minv2}).

%
%

\begin{table}[t]
\begin{center}
\begin{tabular}{|c|c|c|c|c|}
\hline
& \multicolumn{4}{c|}{}\\[-0.2cm]
$m_{(n)}$ [MeV] & \multicolumn{4}{c|}{${\cal M}_{\rm GTF} (m_{(n)})$} \\[0.2cm]
\hline
& & & & \\[-0.2cm]
& $^{76}$Ge & $^{82}$Se & $^{100}$Mo & $^{116}$Cd \\[0.2cm]
\hline
 $\le 1$ & $4.33$ &$4.03 $&$4.86$&$ 3.29$\\
 $10$ & $4.34$ &$4.04 $&$4.81$&$ 3.29$\\
 $10^{2}$ & $3.08$ &$2.82 $&$3.31$&$ 2.18$\\
 $10^{3}$ & $1.40\times 10^{-1}$ &$1.25\times 10^{-1} $&$1.60\times
10^{-1}$&$ 9.34\times 10^{-2}$\\
 $10^{4}$ & $1.39\times 10^{-3}$ &$1.24\times 10^{-3} $&$1.60\times
10^{-3}$&$ 9.26\times 10^{-4}$\\ 
 $10^{5}$ & $1.39\times 10^{-5}$ &$1.24\times 10^{-5} $&$1.60\times
10^{-5}$&$ 9.26\times 10^{-6}$\\ 
 $10^{6}$ & $1.39\times 10^{-7}$ &$1.24\times 10^{-7} $&$1.60\times
10^{-7}$&$ 9.26\times 10^{-8}$\\ 
 $10^{7}$ & $1.39\times 10^{-9}$ &$1.24\times 10^{-9} $&$1.60\times
10^{-9}$&$ 9.26\times 10^{-10}$\\[0.1cm] 
\hline
\hline
& \multicolumn{4}{c|}{}\\[-0.2cm]
$m_{(n)}$ [MeV] & \multicolumn{4}{c|}{${\cal M}_{\rm GTF} (m_{(n)})$}\\[0.2cm]
\hline
& & & & \\[-0.2cm]
& $^{128}$Te & $^{130}$Te & $^{136}$Xe & $^{150}$Nd \\[0.2cm]
\hline
 $\le 1$ & $4.50$&$3.89$&$1.83$&$5.30$\\
 $10$ & $4.52$&$3.91$&$1.88$&$5.45$\\
 $10^{2}$ & $3.19$&$2.79$&$1.48$&$4.24$\\
 $10^{3}$ & $1.46\times 10^{-1}$&$1.29\times 10^{-1}$&$7.07\times
10^{-2}$&$2.02\times 10^{-1}$\\  
 $10^{4}$ & $1.46\times 10^{-3}$&$1.28\times 10^{-3}$&$7.04\times
10^{-4}$&$2.02\times 10^{-3}$\\ 
 $10^{5}$ & $1.46\times 10^{-5}$&$1.28\times 10^{-5}$&$7.05\times
10^{-6}$&$2.02\times 10^{-5}$\\ 
 $10^{6}$ & $1.46\times 10^{-7}$&$1.28\times 10^{-7}$&$7.05\times
10^{-8}$&$2.02\times 10^{-7}$\\ 
 $10^{7}$ & $1.46\times 10^{-9}$&$1.28\times 10^{-9}$&$7.05\times
10^{-10}$&$2.02\times 10^{-9}$\\[0.1cm] 
\hline
\end{tabular}
\end{center}
\caption{\em QRPA estimates of the relevant combination of nuclear
matrix elements, ${\cal M}_{\rm GTF} = {\cal M}_{\rm GT}-{\cal M}_{\rm
F}$, as a function of the KK neutrino mass $m_{(n)}$.}\label{Tab1}
\end{table}

\begin{table}[t]
{\small 
\begin{center}
\begin{tabular}{|l|r|c|c|c|c|c|c|c|c|r|}
\hline
& \multicolumn{9}{c|}{} & \\[-0.2cm]
$1/a$  & \multicolumn{9}{c|}{$\langle m \rangle$~[eV]} & 
                             $\langle m^{-1} \rangle$\\[0.2cm] 
\hline
 & & & & & & & & & & \\[-0.2cm]
[GeV] & SA & $^{76}$Ge & $^{82}$Se & $^{100}$Mo & $^{116}$Cd & $^{128}$Te & 
$^{130}$Te & $^{136}$Xe & $^{150}$Nd & [TeV$^{-1}$]\\[0.2cm]
\hline
0.05  &--0.062 & 0.009 & 0.010 & 0.016 & 0.012 & 0.009 & 0.008 & --0.004  &
                                               --0.004 & 6.2$\times 10^{-6}$\\ 
0.1   &--0.012 & 0.052 & 0.054 & 0.061 & 0.062 & 0.052 & 0.050 & 0.025 & 
                                               0.026 & --3.6$\times 10^{-6}$\\ 
0.2   & 0.208 & 0.096  & 0.100 & 0.109 & 0.114 & 0.097 & 0.094 & 0.058 & 
                                              0.061 & --1.3$\times 10^{-5}$\\ 
0.3   & 0.307 & 0.123   & 0.128 & 0.136 & 0.143 & 0.124 & 0.121 & 0.082 & 
                                              0.086 & --1.2$\times 10^{-5}$\\ 
1     & 0.457 & 0.271 & 0.275 & 0.280 & 0.287 & 0.272 & 0.269 & 0.241 & 
                                           0.243 & --5.7$\times 10^{-6}$\\ 
10    & 0.516 & 0.493 & 0.493 & 0.494 & 0.495 & 0.493 & 0.493 & 0.489 & 0.489 
                                                  & --6.6$\times 10^{-7}$\\ 
\hline
10$^2$& 0.515 & \multicolumn{8}{c|}{\hspace{-1.5cm}0.513} 
& --6.7$\times 10^{-8}$\\ 
\hline
10$^3$& \multicolumn{9}{c|}{0.535} & --6.7$\times 10^{-8}$ \\ 
10$^4$& \multicolumn{9}{c|}{0.066} & --6.9$\times 10^{-10}$\\ 
10$^{10}$ & \multicolumn{9}{c|}{$\stackrel{<}{{}_\sim} 10^{-6}$} & 0.~~~~ \\
\hline
\end{tabular}  
\end{center} }
\caption{\em Numerical estimates of $\langle m \rangle$ for different
nuclei in a 5-dimensional brane-shifted model, with $m=10$~eV, $1/R =
300$~eV, $\varepsilon = 1/(4R)$, $\phi_h = -\pi/4$ and $M_F = 1$~TeV.
The first column exhibits the numerical values for $\langle m \rangle$
in the Step Approximation (SA) for the nuclear matrix elements, while
the last column shows the results for the inverse effective neutrino
mass $\langle m^{-1} \rangle$.}\label{Tab2}
\end{table}

In Table~\ref{Tab1}, we present numerical values for the difference of
the nuclear  matrix elements,  ${\cal  M}_{\rm  GTF} =  {\cal  M}_{\rm
GT}-{\cal   M}_{\rm F}$,  as   a function   of the   KK  neutrino mass
$m_{(n)}$.  Our estimates   are     obtained within  the     so-called
Quasi-particle         Random           Phase            Approximation
(QRPA)~\cite{Muto:cd,Staudt:qi}.   Here,   we  should note   that  the
numerical values for the nuclear  matrix element of $^{100}$Mo exhibit
some  instability  due     to  its  sensitive     dependence   on  the
particle--particle coupling  $g_{PP}$ within the  context of the QRPA.
In addition,  we should  remark that in  our  numerical  evaluation of
$\langle m \rangle$, the  nuclear matrix elements ${\cal M}_{\rm GTF}$
have been interpolated between the values given in Table~\ref{Tab1}.

In Table~\ref{Tab2},  we  show   numerical values for   the  effective
Majorana-neutrino  mass $\langle m \rangle$   as derived for different
nuclei in a 5-dimensional brane-shifted model,  with $m=10$~eV, $1/R =
300$~eV, $\varepsilon = 1/(4R)$, $\phi_h = -\pi/4$  and $M_F = 1$~TeV.
In addition, we have varied discretely  the brane-shifting scale $1/a$
from 0.05~GeV up to  values much larger  than $M_F$.  The first column
in  Table~\ref{Tab2} give the predictions obtained  in the  SA for the
nuclear matrix elements.     The SA   is   closely related    to   our
approximative method followed above, leading to  results that are in a
very good  agreement with~(\ref{mnueff}).  Remarkably enough, even the
change of sign of $\langle m\rangle_{\rm SA}$ at $1/a \approx 0.1$~GeV
in   Table~\ref{Tab2} can be    determined sufficiently accurately  by
analyzing the multiplicative expression   $\pi/2 - {\rm   Si}(2a q_F)$
in~(\ref{mnueff}), which oscillates around $\pi/2$~\cite{AS}, for $1/a
\stackrel{<}{{}_\sim} 0.1$~GeV. Analogous remarks can  be made for the
inverse   effective      neutrino    mass   $\langle    m^{-1}\rangle$
in~(\ref{minv2}).

As can be seen from Table~\ref{Tab2}, the deviation between the SA and
the  one based  on  the general  formula~(\ref{effMajmass}) is  rather
significant if $a$ is close  to $1/q_F$ due to the non-trivial nuclear
matrix element effects mentioned  above and due to heavier KK-neutrino
effects  coming from  $\langle m^{-1}\rangle$.   However,  for smaller
values  of  $a$,  i.e.\  for  $a\stackrel{<}{{}_\sim}  1/(3q_F)$,  the
agreement between the  effective neutrino mass computed in  the SA and
the  general  formula~(\ref{effMajmass})  is  fairly  good.   In  this
kinematic  regime,  the   inverse  effective  neutrino  mass  $\langle
m^{-1}\rangle$ becomes  rather suppressed according  to our discussion
in~(\ref{minv2}).  Our  numerical  estimates  in the  last  column  of
Table~\ref{Tab2} offer  firm support  of this last  observation. Thus,
the  main  contribution to  $\langle  m  \rangle$  originates from  KK
neutrinos   much  lighter  than   $q_F$.   Consequently,   within  the
5-dimensional brane-shifted  model, we have  numerically established a
sizeable  value for $\langle  m \rangle$  in the  presently explorable
range 0.05--0.84~eV. Finally, for very  small values of $a$, i.e.\ for
$a \ll  1/M_F$, we recover the  undetectably small result~(\ref{mnu1})
for the unshifted brane $a=0$.

\subsection{$\langle m \rangle$ and the neutrino mass scale}

Apart from  explaining the  recent excess in  $0\nu\beta\beta$ decays,
the 5-dimensional model  with a small  but non-vanishing shifted brane
exhibits      another   very  important     property.   The  effective
Majorana-neutrino mass $\langle  m \rangle$ can  be  several orders of
magnitude larger than  the light neutrino   mass $m_\nu$, for  certain
choices of the parameters  $\varepsilon$ and $\phi_h$.  To  understand
this   phenomenon,   let      us  first    consider   the   eigenvalue
equation~(\ref{eigen}) for $\lambda = m_\nu$, written in the form:
\begin{equation}
  \label{B2approx}
m_\nu\: +\: \sum\limits_{n=-\infty}^\infty\,\frac{m^{(n) 2}}{
 \varepsilon\: +\: \frac{n}{R}\: -\: m_\nu  }\ =\ 0\; .
\end{equation}
Notice  that  (\ref{B2approx})   constitutes  an  excellent  and  very
practical  approximation of the  neutrino-mass--mixing sum  rule, when
the small $m_\nu$-dependence in the  infinite sum over the KK neutrino
states  is  neglected  and  the  approximate  formulae~(\ref{mn})  and
(\ref{Nnfinal}) for the KK masses $m_{(n)}$ and mixing-matrix elements
$B_{e,n}$, along  with $B_{e\nu} = 1$,  are substituted in~(\ref{B2}).
Then, the infinite  sum over KK neutrino states  can be performed with
the  help of~(\ref{transII})  for rational  values of  $a$ in  $\pi R$
units.  Especially  for $a = \pi  R/q$ with $q$ being  an integer much
larger than  1, i.e.\ for  $1/M_F \ll a  \stackrel{<}{{}_\sim} 1/q_F$,
the light neutrino mass $m_\nu$ is given by
\begin{equation}
  \label{mnu}
m_\nu \ \approx\ -\, \pi m^2 R\ \Big[ \cos^2 \phi_h\, 
\cot (\pi R\,\varepsilon )\ +\ \frac{1}{2}\sin(2\phi_h)\, \Big]\, .
\end{equation}
It is now easy to see that the light neutrino mass $m_\nu$ can be very
suppressed  for  specific values  of $\phi_h$  and $\varepsilon$.  For
instance, one obvious choice   would  be $\phi_h \approx -\pi/4$   and
$\varepsilon \approx 1/(4R)$.   On the    other hand, the    effective
neutrino mass $\langle m \rangle_{\rm SA}$ is determined by the second
sine-dependent term in~(\ref{mnu})  (cf.\  (\ref{mnueff})),  which  is
induced  by  brane-shifting  effects.    Unlike the  suppressed  light
neutrino mass $m_\nu$, the effective neutrino mass $\langle m \rangle$
can be sizeable in the  observable range 0.05--0.84~eV.  This loss  of
correlation between the quantities  $\langle m \rangle$ and $m_\nu$ is
a  rather  unique feature   of  our   higher-dimensional brane-shifted
scenario.  As  we   will  discuss  in   the next  section,  the  above
de-correlation property plays a key  r\^ole  in our model-building  of
5-dimensional brane-shifted scenarios that  could explain the neutrino
oscillation data.

\setcounter{equation}{0}
\section{Atmospheric and solar neutrino data} 

Atmospheric  and solar  neutrino data~\cite{SK,SNO,AF},  together with
information   from   laboratory  experiments,   such   as  the   CHOOZ
experiment~\cite{CHOOZ},    are    very    crucial   for    a    given
higher-dimensional  singlet-neutrino model to  qualify as  viable.  In
particular,  the  latest SNO  results~\cite{SNO}  appear to  disfavour
large components  of sterile neutrinos, indicating  a preference among
the different solutions to  the solar and atmospheric neutrino puzzles
for    those    involving    transitions   between    almost    active
neutrinos.\footnote{A recent study \cite{HS} seems to suggest that the
active neutrino component in the solar neutrinos has to be larger than
86\% at 1~$\sigma$ CL. A loophole may exist for atmospheric neutrinos,
see \cite{PSW}.   }  To account  for this experimental  indication, we
assume    that   the    compactification   scale    $1/R$    and   the
lepton-number-violating bulk  parameter $\varepsilon$ are  much larger
than the KK Dirac mass terms $m^{(n)}$ in~(\ref{mk0}).

In the  following,     we shall   explicitly   demonstrate  that   our
5-dimensional brane-shifted model with only  one bulk neutrino is able
to fully explain the neutrino  oscillation data. Specifically, we will
show that the preferred solar Large Mixing Angle (LMA) and atmospheric
solutions,    which   both   require   large   $\nu_e$-$\nu_\mu$   and
$\nu_\mu$-$\nu_\tau$ mixings, can be realized within our 5-dimensional
model.  These particular    solutions   are  allowed, only  if     the
differences of the squares  of  the light  neutrino masses lie  in the
ranges:
\begin{equation}
  \label{nudata}
1.8\times
10^{-3}\ <\ \Delta m^2_{\rm atm}~[{\rm eV}^2]\ <\ 4.0\times
10^{-3}\,,\qquad
2.0\times
10^{-5}\ <\ \Delta m^2_\odot~[{\rm eV}^2]\ <\ 2.0\times 10^{-4}\,,
\end{equation}
with $\Delta m^2_{\rm  atm} = m^2_{\nu_3}  - m^2_{\nu_2}$ and  $\Delta
m^2_\odot = m^2_{\nu_2}   -  m^2_{\nu_1}$.   According to  the   usual
conventions,  the physical     light   neutrino masses    $m_{\nu_1}$,
$m_{\nu_2}$ and   $m_{\nu_3}$ are labelled  in increasing hierarchical
order, i.e.\ $m_{\nu_1} \le m_{\nu_2} \le m_{\nu_3}$.

To start  with, let us consider  the weak basis  in which  the charged
lepton mass  matrix   is  diagonal.   Then,   in  the three-generation
brane-shifted  model, the KK-Dirac   Yukawa  terms  are given  by  the
3-vectors
\begin{equation}
  \label{mn3g}
{\bf m}^{(n)} \ =\ \left(\! \begin{array}{c}
m^e\, \cos ( \frac{n a}{R}\: -\: \phi_e ) \\
m^\mu\, \cos ( \frac{n a}{R}\: -\: \phi_\mu ) \\
m^\tau\, \cos ( \frac{n a}{R}\: -\: \phi_\tau ) \end{array}\!\right)\,,
\end{equation}
where 
\begin{equation}
  \label{ml3g}
m^l\ =\ \frac{v}{\sqrt{2}}\, \sqrt{ (\bar{h}^l_1)^2\: +\: 
(\bar{h}^l_2)^2\, }\;, \qquad \phi_l \ =\ \tan^{-1} \Bigg(
\frac{\bar{h}^l_2}{\bar{h}^l_1}\Bigg)\ +\ \frac{k_0 a}{R}\ ,
\end{equation}
with $l=e,\mu ,\tau$.   Given our assumption that $\varepsilon\,,  1/R
\gg m^l$, the KK   neutrinos can now   be integrated out.  Analogously
with~(\ref{eigen}), the  effective light  neutrino mass matrix  ${\cal
M}^\nu$ can be computed by
\begin{equation}
  \label{Mnueff1}
{\cal M}^\nu\ =\  -\, \sum\limits_{n=-\infty}^{+\infty}\
\frac{{\bf m}^{(n)}\,{\bf m}^{(n)\, T}}{\frac{n}{R}\: +\: \varepsilon}\ .
\end{equation}
Following the same   line of steps as   in Appendix~A, one is able  to
analytically carry  out  the infinite sum in~(\ref{Mnueff1})   for the
phenomenologically interesting  case of $a = \pi  R/q$, with $q$ being
an  integer much larger than  1. In  this limit,  we  obtain the novel
trigonometric mass texture:
\begin{equation}
  \label{Mnueff}
{\cal M}^\nu_{ll'}\ =\ -\, 
\pi R\, m^l m^{l'}\, \bigg[\, \cos\phi_l\,\cos\phi_{l'}\,\cot (\pi
R\varepsilon)\ +\ \frac{1}{2}\, \sin (\phi_l +\phi_{l'} )\, \bigg]\, ,
\end{equation}
with   $l,l'    = e, \mu,     \tau$.   The  effective   neutrino  mass
matrix~(\ref{Mnueff}) consists of  two terms: (i)~the cosine-dependent
term that arises from the   lepton-number-violating bulk mass $M$  (or
equivalently $\varepsilon$) and (ii)~the  sine-dependent term which is
due  to lepton-number violation  in the effective Yukawa couplings and
is caused  by slightly  shifting  the brane  from  the  orbifold fixed
points.  The occurrence of  the   second brane-shifting mass  term  is
always ensured as long as $a \gg 1/M_F$.  Without the presence of this
brane-shifting-induced   term,   the      effective  neutrino     mass
matrix~(\ref{Mnueff}) is of rank~1, leading to two massless neutrinos.
This last fact is  very undesirable, as it would  be very difficult to
explain both solar   and   atmospheric neutrino  data  with  only  one
non-trivial difference  of neutrino masses in the frequently-discussed
scenario without brane shifting.

As has been discussed  in Section~4, however, even  a small amount  of
brane  shifting   may  induce sizeable lepton-number-violating  Yukawa
interactions.   The latter   generate brane-shifting mass   terms that
break  the    rank-1 structure    of  the   effective   neutrino  mass
matrix~${\cal M}^\nu$.  The resulting ${\cal M}^\nu$ in~(\ref{Mnueff})
exhibits a novel trigonometric structure that can predict hierarchical
neutrinos    with  large  $\nu_\mu$--$\nu_\tau$ and $\nu_\mu$--$\nu_e$
mixings to explain the atmospheric and solar neutrino anomalies, along
with  a  small $\nu_e$--$\nu_\tau$  mixing  as required  by  the CHOOZ
experiment~\cite{CHOOZ}.  At  this  point, it is   important to stress
that the effective   neutrino  mass $\langle m\rangle$  entering   the
$0\nu\beta\beta$-decay  amplitude gets    fully  decoupled  from   the
neutrino-mass matrix  element ${\cal M}^\nu_{ee}$.   According to  our
discussions  in Section~4 (cf.~(\ref{mnueff})), the effective neutrino
mass for the three-generation case is given by
\begin{equation}
  \label{meff3}
\langle m \rangle\ \approx - \frac{1}{2}\ \sin (2\phi_e)\, \pi (m^e)^2\, R\,
\ \neq\ {\cal M}^\nu_{ee}\ .
\end{equation}
It is important to recall   again that unlike ${\cal M}^\nu_{ee}$,  KK
neutrinos  heavier than   the Fermi   nuclear  momentum $q_F$  do  not
contribute significantly to $\langle m\rangle$, leading to the loss of
correlation between  $\langle m\rangle$ and  ${\cal M}^\nu_{ee}$.  The
latter is  a distinctive  feature of the  KK-neutrino  dynamics.  This
de-corellation  between  $\langle  m\rangle$  and ${\cal  M}^\nu_{ee}$
permit us  to consider the interesting case  $|\langle m  \rangle| \gg
|{\cal   M}^\nu_{ll'}|$, for   all  $l,l'  =  e,\mu,   \tau$.  Such  a
realization enables us to  accommodate  a sizeable positive  signal of
$0\nu\beta\beta$ decays together with the present neutrino oscillation
data.

To realize the   aforementioned  hierarchy $|\langle  m   \rangle| \gg
|{\cal M}^\nu_{ll'}|$, we assume that all phases $\phi_l$ are close to
$-\pi/4$.  For concreteness, we adopt the following scheme of phases:
\begin{equation}
  \label{phases}
\phi_l \ =\ -\, \frac{\pi}{4}\: +\: \delta_l\; ,\qquad 
\pi R \varepsilon\ =\ \frac{\pi}{4}\: -\: \delta_\varepsilon\ .
\end{equation}
where $\delta_l,\ \delta_\varepsilon \ll 1$.  Our choice of phases
has been motivated by the fact that the above-described de-correlation
between   $\langle m \rangle$ and   ${\cal  M}^\nu_{ee}$ becomes fully
operative in this  case.  To  implement  the CHOOZ constraint  in  our
model-building, we  require    that  ${\cal M}^\nu_{e\tau}    =  {\cal
M}^\nu_{\tau e} = 0$.  This last constraint implies that
\begin{equation}
2\delta_\varepsilon\ =\ -\delta_e - \delta_\tau\; .
\end{equation}
Moreover, without loss of generality  within our phase scheme, we  may
take   $\delta_\mu  = 0$.     Under  these  assumptions,  the  light
neutrino-mass matrix takes on the simple form
\begin{equation}
  \label{mtexture}
{\cal M}^\nu\ = \ \frac{\pi R}{2}\, \left(\! \begin{array}{ccc}
m^{e\, 2}\, (\delta_\tau -\delta_e)  & m^e m^\mu\, \delta_\tau & 0 \\
m^\mu m^e\, \delta_\tau & m^{\mu\, 2}\, (\delta_e + \delta_\tau) & 
m^\mu m^\tau\, \delta_e \\
0 & m^\tau m^\mu\, \delta_e & m^{\tau\, 2}\, 
( \delta_e - \delta_\tau) \end{array} \!\right)\; .
\end{equation}
Let us now consider the following numerical example:
\begin{eqnarray}
  \label{numerics}
\delta_\tau &=& \delta\,,\qquad \delta_e\ =\ 2\delta\,,\nonumber\\
\frac{m^\mu}{m^e}  
&\approx& 1.468\ ,\qquad 
\frac{m^\tau}{m^e} 
\ \approx \ 2.542\, . 
\end{eqnarray}
This leads to the neutrino mass matrix:
\begin{equation}
  \label{mnuexample}
{\cal M}^\nu\ = \ \delta\ \frac{\pi m^{e\, 2} R}{2}\, 
\left(\! \begin{array}{ccc}
-1  & 1.47 & 0 \\
1.47 & 6.46 & 7.46 \\
0 & 7.46 & 6.46 \end{array} \!\right)\; .
\end{equation}
Notice that  all elements of  the neutrino-mass matrix  ${\cal M}^\nu$
in~(\ref{mnuexample}) can be suppressed  by choosing a small value for
the factorizable parameter $\delta$.   In our numerical example, the
neutrino  mass matrix~(\ref{mnuexample})  can be  diagonalized through
$\nu_\mu$-$\nu_\tau$  and  $\nu_e$-$\nu_\mu$  mixing angles  close  to
$\pi/4$, whereas  the $\nu_e$-$\nu_\tau$ mixing angle  is small, below
0.1.  In addition, its mass-eigenvalues are approximately given by
\begin{equation}
({\cal M}^\nu)_{\rm diag}\ \approx\  \delta\,\pi m^{e\, 2} R\, 
\Big( 0\,,\ 1\,,\ 7\,\Big)\; .
\end{equation}
Assuming  that $m^e  =  10$~eV and $1/R  =  300$~eV  for a  successful
interpretation of the recent  excess in $0\nu\beta\beta$ decays,  then
it should be  $\delta = (6$--$9)\times  10^{-3}$ to  accommodate the
neutrino oscillation data through the  LMA solution. In particular, we
obtain the neutrino-mass differences:
\begin{equation}
  \label{mnudiffs}
\Delta m^2_{\rm atm}\ \approx\ (2-4)\times 10^{-3}~{\rm eV}^2\,,\qquad
\Delta m^2_{\odot}\ \approx\ (4-8)\times 10^{-5}~{\rm eV}^2
\end{equation}
These  results are   fully  compatible with   the  currently preferred
atmospheric and solar LMA solutions to the neutrino anomalies.

In our demonstrative analysis carried out in this section, we have not
attempted  to fit the    results of the  Liquid  Scintillator Neutrino
Detector (LSND) as well~\cite{LSND}.   In principle, our brane-shifted
5-dimensional  models  are capable of   accommodating the LSND results
through active-sterile neutrino  transitions.  In this  case, however,
the lowest-lying KK singlet neutrinos should be relatively light. As a
result, they    cannot be integrated    out  from the  light  neutrino
spectrum, thereby  leading    to  a  much   more  involved   effective
neutrino-mass  matrix.    A complete  study of   this issue, including
possible    constraints    from      the    cooling    of    supernova
SN1987A~\cite{LRRR,SN1987A}, is beyond the  scope of the present paper
and may be given elsewhere.

\setcounter{equation}{0}
\section{Conclusions} 

We  have  studied  the  model-building constraints  derived  from  the
requirement  that KK singlet  neutrinos in  theories with  large extra
dimensions can  give rise to a  sizeable $0\nu\beta\beta$-decay signal
to  the level  of~0.4~eV  reported recently.   Our  analysis has  been
focused on  5-dimensional $S^1/Z_2$  orbifold models with  one sterile
(singlet) neutrino in the bulk,  while the SM fields are considered to
be  localized on  a  3-brane.   In our  model-building,  we have  also
allowed the 3-brane to be  displaced from the $S^1/Z_2$ orbifold fixed
points.   Within this  minimal 5-dimensional  brane-shifted framework,
lepton-number  violation  can   be  introduced  through  Majorana-like
bilinears,  which  may  or  may  not arise  from  the  Scherk--Schwarz
mechanism,  and   through  lepton-number-violating  Yukawa  couplings.
However, lepton-number-violating  Yukawa couplings can  be admitted in
the theory, only if the 3-brane is shifted from the $S^1/Z_2$ orbifold
fixed  points.   Apart   from  a  possible  stringy  origin~\cite{GP},
brane-shifting might also be regarded  as an effective result owing to
a  non-trivial 5-dimensional profile  of the  Higgs particle~\cite{RS}
and/or other SM fields~\cite{AHS,KKS} that live in different locations
of a 3-brane  with non-zero thickness which is centered  at one of the
$S^1/Z_2$ orbifold fixed points.

One   major difficulty of the    higher-dimensional theories is  their
generic   prediction  of a   KK  neutrino  spectrum  of  approximately
degenerate states with opposite  CP parities that lead to  exceedingly
suppressed values for the  effective Majorana-neutrino mass $\langle m
\rangle$.  Nevertheless, we have  shown that within the  5-dimensional
brane-shifted framework, the  KK neutrinos can  couple  to the $W^\pm$
bosons with unequal  strength, thus avoiding the  disastrous CP-parity
cancellations in the  $0\nu\beta\beta$-decay amplitude. In particular,
the     brane-shifting parameter  $a$   can   be  determined from  the
requirement that the effective Majorana mass $\langle m \rangle$ is in
the observable range~\cite{HMexp}: 0.05--0.84~eV. In this way, we have
found  that  $1/a$ has  to be  larger  than the  typical Fermi nuclear
momentum $q_F  = 100~{\rm  MeV}$  and much  smaller  than the  quantum
gravity     scale    $M_F$,   or     equivalently    $1/M_F    \ll   a
\stackrel{<}{{}_\sim} 1/q_F$.

An important  prediction of our  5-dimensional  brane-shifted model is
that the effective Majorana-neutrino mass  $\langle m \rangle$ and the
scale of light neutrino masses   can be completely de-correlated   for
certain    natural  choices  of     the   Majorana-like bilinear  term
$\varepsilon$ and the  original 5-dimensional Yukawa couplings $h^l_1$
and  $h^l_2$ in~(\ref{Leff}).   For example,  if  $\varepsilon \approx
1/(4R)$ and  $h^l_1 \approx -  h^l_2$, we obtain light-neutrino masses
that   can be several  orders  of  magnitude smaller  than $\langle  m
\rangle$.  Nevertheless, it is worth  mentioning  that if future  data
did not substantiate the  presently reported $0\nu\beta\beta$  excess,
the above model-building conditions would then  need be modified. Such
a possible modification  would not jeopardize  though the viability of
our  brane-shifted  scenario.  Indeed,   if  the  upper limit  on  the
effective neutrino mass became even lower and  lower, this would imply
that the above decorrelation property is less and less necessary.

Another important prediction of  the 5-dimensional brane-shifted model
with  {\em  only  one}  bulk  sterile neutrino   is that the  emerging
effective light-neutrino mass matrix does no longer possess the rank-1
form, as opposed to the brane-unshifted $a=0$ case.   As we have shown
in Section~5, the above   properties of the brane-shifted models   are
sufficient  to explain, even with only  one neutrino  in the bulk, the
present  solar and atmospheric neutrino data  by means of oscillations
of hierarchical neutrinos   with large $\nu_e$-$\nu_\mu$  and  maximal
$\nu_\mu$-$\nu_\tau$ mixings.   In particular, neutrino-mass  textures
can be constructed that utilize the  currently preferred LMA solution,
where the  $\nu_e$-$\nu_\tau$ mixing is  small  in agreement  with the
CHOOZ experiment.

Although a  sizeable  $0\nu\beta\beta$-decay signal  can be  predicted
within our   brane-shifted 5-dimensional models,  the  above-described
de-correlation property between $\langle   m \rangle$ and   the actual
light  neutrino masses suggests, however, that  it  is rather unlikely
that such a signal be accompanied by a corresponding signal in Tritium
beta-decay  experiments. For example, the KATRIN project~\cite{KATRIN}
has a  sensitivity to active   neutrino masses larger  than 0.35~eV at
95\%~CL, and  so it can  only probe the   existence of light neutrinos
much  heavier than   those  considered in   our  5-dimensional models.
Finally, the brane-shifted models under study  also have the potential
to  accommodate the LSND  results by virtue of active-sterile neutrino
oscillations.  In this case,  the lowest-lying KK-neutrino states will
contribute to the effective light neutrino-mass matrix, giving rise to
more  involved mass  textures.   In  this context,   it would be  very
interesting   to  investigate    the   question   whether   a   simple
higher-dimensional model  accounting  for  all the  observed  neutrino
anomalies  can  be established.   We  plan to address this interesting
question in the near future.

\subsection*{Acknowledgements} We thank Martin Hirsch for discussions 
on QRPA computations and Antonio Delgado for comments on the geometric
breaking of lepton number violation in higher-dimensional theories.

\newpage

\def\theequation{\Alph{section}.\arabic{equation}}
\begin{appendix}

\setcounter{equation}{0}
\section{Eigenvalue equation}

Starting  from~(\ref{eigen}), we will  derive here  the transcendental
eigenvalue  equation~(\ref{transII}),   for  the  simplest   class  of
brane-shiftings with $a = \pi R/q$,  where $r=1$ and $q$ is an integer
larger   than    1,   i.e.\   $q\ge   2$.     Then,   the   eigenvalue
equation~(\ref{eigen}) can be equivalently written as
\begin{eqnarray}
  \label{sum1}
\lambda &=& \sum_{l=0}^{q-1}\, \sum_{k=-\infty}^\infty\, 
\frac{m^{(qk+l)\, 2}}{\lambda - \varepsilon - \frac{qk+l}{R}}\
=\ \sum_{l=0}^{q-1}\ m^{(l)\, 2}\, \sum_{k=-\infty}^\infty\, 
\frac{1}{\lambda - \varepsilon - \frac{qk+l}{R}}\ ,
\end{eqnarray}
where   we  have  used   the    periodicity  property  $(m^{(l)})^2  =
(m^{(qk+l)})^2$ in the  second  step of~(\ref{sum1}).  In~fact,  it is
this last periodicity property of the KK-Yukawa  terms that we wish to
exploit  here     to  carry out     analytically   the  infinite  sums
in~(\ref{sum1}), which  has    forced us to   introduce  the technical
constraint~(\ref{rat}), namely that $a/(\pi R)$  is a rational number.
Now,   the individual   $l$-dependent     infinite  sums   over    $k$
in~(\ref{sum1}) can be  performed independently, using complex contour
integration techniques. In this way, we obtain
\begin{equation}
  \label{sum2}
\lambda \ =\ \frac{1}{q}\, \pi m^2 R\ \sum_{l=0}^{q-1}\ 
\cos^2\bigg( \phi_h\: -\: \frac{l\pi}{q}\,\bigg)\, 
\cot \bigg[\,\frac{1}{q}\,\pi R\, (\lambda - \varepsilon )\: 
-\: \frac{l\pi}{q}\,\bigg] \; .
\end{equation} 

Our next task is to carry out the  summation over $l$ in~(\ref{sum2}).
For this purpose, we express the RHS of~(\ref{sum2}) entirely in terms
of sine and cosine functions by factoring out the common divisor, i.e.
\begin{eqnarray}
  \label{prod1}
\lambda &=& \frac{\pi m^2 R}{q\, \prod\limits_{l=0}^{q-1}\, \sin \Big(
\frac{\theta}{q} - \frac{l\pi}{q}\Big)}\ \sum_{l=0}^{q-1}\ 
\cos^2\bigg( \phi_h\: -\: \frac{l\pi}{q}\,\bigg)\, 
\cos \bigg(\, \frac{\theta}{q}\: -\: \frac{l\pi}{q}\,\bigg)\,
\prod\limits_{\stackrel{m=0}{{}_{(m\ne l)}}}^{q-1}
\sin \bigg(\, \frac{\theta}{q}\: -\: \frac{m\pi}{q}\,\bigg)\,,\quad
\end{eqnarray}
with $\theta = \pi R\,  (\lambda - \varepsilon )$. To further evaluate
(\ref{prod1}),     we    exploit    the     following    trigonometric
identities:\footnote{The proof  of these identities  is rather lengthy
and relies on the particular properties of the $q$-roots of the unity,
i.e.\ the roots of the equation  $z^q = 1$.  Specifically, we used the
basic property of  the unit roots that their sum and  the sum of their
products are zero, while their total product is $(-1)^{q-1}$.}
\begin{eqnarray}
  \label{id1}
\prod\limits_{l=0}^{q-1}\, \sin \bigg(\,
\frac{\theta}{q} - \frac{l\pi}{q}\,\bigg) & =& \frac{(-1)^{q-1}}{2^{q-1}}\
\sin\theta\;, \\
  \label{id2}
\sum_{l=0}^{q-1}\ 
\cos \bigg(\, \frac{\theta}{q}\: -\: \frac{l\pi}{q}\,\bigg)\,
\prod\limits_{\stackrel{m=0}{{}_{(m\ne l)}}}^{q-1}
\sin \bigg(\, \frac{\theta}{q}\: -\: \frac{m\pi}{q}\,\bigg) &=& 
\frac{(-1)^{q-1}}{2^{q-1}}\ q\, \cos\theta\;,\\
  \label{id3}
\sum_{l=0}^{q-1}\ 
\cos\bigg( 2\phi_h\: -\: \frac{2l\pi}{q}\,\bigg)\, 
\cos \bigg(\, \frac{\theta}{q}\: -\: \frac{l\pi}{q}\,\bigg)\,
\prod\limits_{\stackrel{m=0}{{}_{(m\ne l)}}}^{q-1}
\sin \bigg(\, \frac{\theta}{q}\: -\: \frac{m\pi}{q}\,\bigg) &=&
\nonumber\\
&&\hspace{-2cm} \frac{(-1)^{q-1}}{2^{q-1}}\ q\ \cos\bigg( 2\phi_h\: +\: 
\frac{q-2}{q}\, \theta\,\bigg)\; .\qquad
\end{eqnarray}
With the  help   of~(\ref{id1})--(\ref{id2}),   we   arrive at     the
transcendental eigenvalue equation
\begin{equation}
  \label{trgen}
\lambda\ =\ \frac{\pi m^2 R}{2}\, \Bigg\{\, 
\cot \Big[\,\pi R\, (\lambda - \varepsilon )\,\Big]\ +\ 
\frac{\cos\Big[ 2\phi_h\: +\: \frac{q-2}{q}\, 
\pi R\, (\lambda - \varepsilon )\,\Big]}
{\sin \Big[\,\pi R\, (\lambda - \varepsilon )\,\Big]}\ \Bigg\}\;.
\end{equation}
If we  replace $q$  with $\pi R/a$  in~(\ref{trgen}), we  arrive after
simple   trigonometric  algebra   at  the   transcendental  eigenvalue
equation~(\ref{transII}).   Although we focused  our attention  on the
simplest  class  with  $a  =  \pi  R/q$, we  should  remark  that  our
methodology described above can apply equally well to the most general
case where the brane-shifting $a$ is any rational number $r/q$ in $\pi
R$ units.

\setcounter{equation}{0}
\section{Sum rules}

In this appendix, we will show  that the KK-neutrino masses determined
by the  roots of~(\ref{transI})  and the mixing-matrix  elements given
in~(\ref{Bln}) satisfy the sum rules~(\ref{B1})  and (\ref{B2}).   For
simplicity, we  consider the  case $a=0$. However,  our considerations
carry over very analogously to the case $a = \pi R/q \ne 0$, where $q$
is an integer larger than 1.

Let us first consider~(\ref{B1}) for $l = l' = e$.  We will then prove
that
\begin{equation}
  \label{AB1}      
|B_{e\nu}|^2\ +\ \lim_{N\to\infty}\ \sum\limits_{n=-N}^{N} |B_{e,n}|^2\ =\ 1\;.
\end{equation}
Our   proof will rely  on Cauchy's   integral  theorem.  Thus, the LHS
of~(\ref{AB1})  can be  expressed in  terms of a  complex  integral as
follows:
\begin{eqnarray}
  \label{AB11}
|B_{e\nu}|^2\ +\ \lim_{N\to\infty}\ \sum\limits_{n=-N}^{N} |B_{e,n}|^2
&=& \frac{1}{2\pi i}\, \lim_{N\to\infty}\ 
\oint_{C_N} dz\ \bigg( \frac{1}{z-m_\nu}\: +\:
\sum\limits_{n=-N}^{N}\ \frac{1}{z-m_{(n)}}\, \bigg)\nonumber\\
&&\times\, \frac{1}{1\: +\: \pi^2 m^2 R^2/\sin^2 [\pi R ( z -\varepsilon)]} 
\nonumber\\[3mm]   
&=&  \frac{1}{2\pi i}\, \lim_{N\to\infty}\ 
\oint_{C_N} dz\ \frac{1}{z\: -\: \pi m^2 R\,\cot [\pi R (z -
\varepsilon )]}\ .
\end{eqnarray}
In deriving the second equality in~(\ref{AB11}),  we have noticed that
for $z$ in the vicinity of the pole, e.g. for  $z \approx m_{(n)}$, it
is
\begin{equation}
z\: -\: \pi m^2 R\,\cot [\pi R (z - \varepsilon )]\ \approx\ 
(z - m_{(n)})\, \bigg\{ \,
1\: +\: \frac{\pi^2 m^2 R^2}{\sin^2 [\pi R (z - \varepsilon)]}\,\bigg\}\ .
\end{equation}
Such a substitution is only valid  under complex integration, provided
there are  no singularities of the complex  function $\cot [\pi R (z -
\varepsilon )]$ on the contour $C_N$. For  this purpose, we choose our
contours to be circles represented in the complex plane as
\begin{equation}
 \label{zN}
z_N \ = \ \frac{(N + \frac{1}{2})\, e^{i\theta}}{R}\
+\ \varepsilon\; . 
\end{equation}
Then, it can be shown  that on the complex  contours $z = z_N$, $|\cot
\pi R (z_N   -\varepsilon  )|$ is  bounded from  above  by a  constant
independent of $N$.  Thus, on  $C_N$ the last integral in~(\ref{AB11})
may be successively computed as
\begin{eqnarray}
  \label{AB12}
\frac{1}{2\pi i}\, \lim_{N\to\infty}\  \oint_{C_N} dz\ 
\frac{1}{z\: -\: \pi m^2 R\,\cot [\pi R (z -
\varepsilon )]}\nonumber\\ 
&&\hspace{-3.5cm} =\ \frac{1}{2\pi i}\, \lim_{N\to\infty}\ 
\int_0^{2\pi} d\theta\ \frac{i (z_N - \varepsilon) }{z_N\: -\: \pi m^2 R\,
\cot [\pi (N + {\textstyle \frac{1}{2}})\, e^{i\theta}]} \nonumber\\[3mm]
&&\hspace{-3.5cm} =\ 1\ +\ \frac{1}{2\pi} \, \lim_{N\to\infty}\ 
\int_0^{2\pi} d\theta\ \frac{\pi m^2 R\,
\cot [\pi (N + {\textstyle \frac{1}{2}})\, e^{i\theta}]
\: -\: \varepsilon}{z_N\: -\: \pi m^2 R\,
\cot[\pi (N + {\textstyle \frac{1}{2}})\, e^{i\theta}]}\ .\qquad
\end{eqnarray}
The second term  in the last  equality of~(\ref{AB12}) vanishes in the
limit $N  \to \infty$ or equivalently  when $z_N$ is taken to infinity
in a  discrete manner as prescribed   by~(\ref{zN}). Thus, the complex
integral in the last   equality  of~(\ref{AB11}) is exactly  1,  which
proves the unitarity sum rule~(\ref{AB1}).

In  the    remainder   of    the  appendix,     we   will    prove the
neutrino-mass-mixing sum rule:
\begin{equation}
  \label{AB2}
B_{e\nu}^2\,m_\nu\ +\ 
\lim_{N\to\infty}\ \sum\limits_{n=-N}^{N} B_{e,n}^2\, m_{(n)}\ =\ 0\;.
\end{equation}
In our proof, we will follow a path very analogous to the one outlined
above  for showing~(\ref{AB1}).  Thus, the   LHS of~(\ref{AB2}) may be
expressed in terms of a complex integral as follows:
\begin{eqnarray}
  \label{AB21}
B_{e\nu}^2\,m_\nu\ +\ 
\lim_{N\to\infty}\ \sum\limits_{n=-N}^{N} B_{e,n}^2\, m_{(n)}
&=&  \frac{1}{2\pi i}\, \lim_{N\to\infty}\ 
\oint_{C_N} dz\ \frac{z}{z\: -\: \pi m^2 R\,\cot [\pi R (z -
\varepsilon )]}\ .\quad
\end{eqnarray}
Evaluating the   complex  integral on    the contours  $C_N$   defined
by~(\ref{zN}) yields
\begin{eqnarray}
  \label{AB22}
\frac{1}{2\pi i}\, \lim_{N\to\infty}\ 
\oint_{C_N} dz\ \frac{z}{z\: -\: \pi m^2 R\,\cot [\pi R (z -
\varepsilon )]}\nonumber\\
&&\hspace{-4cm}=\ \frac{1}{2\pi i}\, \lim_{N\to\infty}\ 
\int_0^{2\pi} d\theta\ \frac{i (z_N\: -\: \varepsilon)\
\pi m^2 R\,
\cot [\pi (N + \frac{1}{2})\, e^{i\theta}]}{z_N\: -\: \pi m^2 R\,
\cot[\pi (N + \frac{1}{2})\, e^{i\theta}]}\nonumber\\[3mm]
&&\hspace{-4cm}=\ \frac{1}{2}\ m^2 R\, \lim_{N\to\infty}\,\bigg\{\,
\int_0^{2\pi} d\theta\ \cot 
[\pi (N + {\textstyle \frac{1}{2}})\, e^{i\theta}] \: +\: 
\ {\cal O}(1/z_N)\, \bigg\}\,.\qquad
\end{eqnarray}
Similar to the second term in the last equality of~(\ref{AB22}), which
goes to zero for $N\to \infty$, the first  term vanishes as well after
integration over $\theta$.   This can  be  readily seen by  exploiting
respectively  the  periodic  and  antisymmetric   properties  of   the
integrand with respect to $\theta$ and its argument:
\begin{eqnarray}
\int_0^{2\pi} d\theta\ \cot 
[\pi (N + {\textstyle \frac{1}{2}})\, e^{i\theta}]& =& 
\int_0^{\pi} d\theta\ \cot 
[\pi (N + {\textstyle \frac{1}{2}})\, e^{i\theta}]\: +\:
\int_\pi^{2\pi} d\theta\ \cot 
[\pi (N + {\textstyle \frac{1}{2}})\, e^{i\theta}]\nonumber\\
&=& \int_0^{\pi} d\theta\ \cot 
[\pi (N + {\textstyle \frac{1}{2}})\, e^{i\theta}]\: +\:
\int_0^{\pi} d\theta\ \cot 
[\pi (N + {\textstyle \frac{1}{2}})\, e^{i(\theta + \pi )}]\nonumber\\
&=& \int_0^{\pi} d\theta\ \cot 
[\pi (N + {\textstyle \frac{1}{2}})\, e^{i\theta}]\: +\:
\int_0^{\pi} d\theta\ \cot 
[-\pi (N + {\textstyle \frac{1}{2}})\, e^{i\theta}]\nonumber\\
& = & 0\, .
\end{eqnarray}
Consequently, the   complex   integral  on  the  RHS   of~(\ref{AB21})
vanishes identically, q.e.d.

\end{appendix}

\end{document}